\documentclass[journal]{IEEEtran}
\IEEEoverridecommandlockouts
\usepackage{cite}
\usepackage{graphicx}
\usepackage{textcomp}
\usepackage{xcolor}
\usepackage{amsmath, amssymb, amsfonts}
\usepackage{bm}
\usepackage{pifont}
\usepackage{soul}
\usepackage{float}

\usepackage{algorithm}
\usepackage{algpseudocode}

\usepackage{multirow}
\usepackage{rotating}
\usepackage{array}
\usepackage{makecell}
\usepackage{tabularx}
\usepackage{enumitem}
\usepackage{url}

\newcolumntype{C}[1]{>{\centering\arraybackslash}m{#1}}
\newcolumntype{Y}{>{\centering\arraybackslash}X}

\newtheorem{theorem}{Theorem}

\usepackage{indentfirst}
\usepackage[caption=false, font=normalsize, labelfont=sf, textfont=sf]{subfig}

\def\BibTeX{{\rm B\kern-.05em{\sc i\kern-.025em b}\kern-.08em
    T\kern-.1667em\lower.7ex\hbox{E}\kern-.125emX}}
\begin{document}

\setlength{\textfloatsep}{5pt}
\setlength{\intextsep}{5pt}
\setlength{\floatsep}{5pt}

\title{Digital Twin-Assisted Data-Driven Optimization for Reliable Edge Caching in Wireless Networks}

\author{
		\IEEEauthorblockN{Zifan Zhang,
            Yuchen Liu,~\IEEEmembership{Member,~IEEE,} 
		Zhiyuan Peng,
		Mingzhe Chen,~\IEEEmembership{Member,~IEEE,} \\
            Dongkuan Xu,~\IEEEmembership{Member,~IEEE,}
            Shuguang Cui,~\IEEEmembership{Fellow,~IEEE} \\
 }

  \thanks{Z. Zhang, Y. Liu, Z. Peng, and D. Xu are with the Department of Computer Science, North Carolina State University, Raleigh, NC, 27695, USA (Email: \{zzhang66, yuchen.liu\}@ncsu.edu). \textit{(Corresponding author: Yuchen Liu.)}}
\thanks{M. Chen is with the Department of Electrical and Computer Engineering and Frost Institute for Data Science and Computing, University of Miami, Coral Gables, FL 33146 USA (Email: \protect\url{ mingzhe.chen@miami.edu)}.} 
\thanks{S. Cui is with the School of Science and Engineering,
The Chinese University of Hong Kong, Shenzhen, China (Email: shuguangcui@cuhk.edu.cn).}
\thanks{This work was supported by the U.S. National Science Foundation under Grants CNS-2312138, CNS-2312139 and CNS-2332834. 
% and NSFC 62293482.
}
\vspace{-1.5em}
	}

\markboth{IEEE Journal on Selected Areas in Communications}%
{Shell \MakeLowercase{\textit{et al.}}: Bare Demo of IEEEtran.cls for IEEE Journals}

\maketitle
% \IEEEpeerreviewmaketitle

\begin{abstract}

Optimizing edge caching is crucial for the advancement of next-generation (nextG) wireless networks, ensuring high-speed and low-latency services for mobile users. Existing data-driven optimization approaches often lack awareness of the distribution of random data variables and focus solely on optimizing cache hit rates, neglecting potential reliability concerns, such as base station overload and unbalanced cache issues. This oversight can result in system crashes and degraded user experience. To bridge this gap, we introduce a novel digital twin-assisted optimization framework, called D-REC, which integrates reinforcement learning (RL) with diverse intervention modules to ensure reliable caching in nextG wireless networks. We first develop a joint vertical and horizontal twinning approach to efficiently create network digital twins, which are then employed by D-REC as RL optimizers and safeguards, providing ample datasets for training and predictive evaluation of our cache replacement policy. By incorporating reliability modules into a constrained Markov decision process, D-REC can adaptively adjust actions, rewards, and states to comply with advantageous constraints, minimizing the risk of network failures. Theoretical analysis demonstrates comparable convergence rates between D-REC and vanilla data-driven methods without compromising caching performance. Extensive experiments validate that D-REC outperforms conventional approaches in cache hit rate and load balancing while effectively enforcing predetermined reliability intervention modules.

\end{abstract}


\IEEEpeerreviewmaketitle

% !TEX root = mainfile.tex

\section{Introduction} \label{sec:intro}

\IEEEPARstart{I}{n} the rapidly evolving scheme of wireless communication, the advent of next-generation (nextG) wireless networks holds the promise of transforming connectivity by enabling unparalleled data transfer speeds and capacities. However, the exponential growth in data transmission poses a formidable challenge in effectively managing network resources to ensure optimal performance. In this context, edge caching, a technique that involves strategically storing frequently accessed data closer to end-users, has emerged as a critical solution to enable more efficient content delivery and low-latency services in nextG wireless networks~\cite{8676308,9417701}.

The role of caching in meeting the escalating demand for high-quality, real-time services within wireless networks cannot be overstated. By caching popular content, such as videos, images, and applications, in close proximity to users, the time and resources required for data retrieval are significantly reduced~\cite{chen2017echo,yang2020learning,wen2020enhancing}. This not only enhances the end-user experience by minimizing latency but also alleviates the strain on network infrastructure, thereby enhancing its overall efficiency. Additionally, caching plays a crucial role in enabling the seamless delivery of data-intensive services, including augmented reality (AR), virtual reality (VR), video streaming, and autonomous vehicles, which are anticipated to become increasingly prevalent in the coming 6G era~\cite{10124955, 8052502, 9552606}.

Traditional caching methods, such as least-recently-used (LRU) and least-frequently-used (LFU), are based on manually engineered heuristics to capture the most common cache access patterns. Their efficacy can be significantly reduced, due to the storage limitation of access points (APs) and the high heterogeneity of user equipment (UE) preference. In the wireless caching optimization problem, certain parameters or variables, such as user demands and preferences, are subject to randomness and variability. This introduces challenges for classical mathematical optimization solutions, as the spatial-temporal shift characteristics make it difficult to guarantee consistent optimality. To tackle this challenge, stochastic optimization and robust optimization methodologies are designed to address scenarios involving random or uncertain data streams and parameters, e.g., in the context of wireless caching. However, stochastic optimization requires knowledge of the exact distribution of random data and variables, while robust optimization, often considered conservative, maximizes worst-case pay-offs and may underperform in practical situations.

Recently, several efforts have been put into data-driven optimization approaches.
\cite{kim2017ultra} investigates ultra-dense edge caching under spatial-temporal demand and network dynamics, where each user can request cached content from multiple small-cell base stations (BSs). A caching algorithm weaving together notions of mean-field game theory and stochastic geometry is proposed to maximize local caching gain and minimize the replicated content caching.
Besides, \cite{wang2017caching} explores a decentralized caching policy where popular contents are cached on UEs and can be shared with other users. To alleviate the heavy BS burden in mobile wireless networks, data-driven recommendation techniques like collaborative filtering and latent factor modeling are incorporated into edge caching.
\cite{ale2019online} proposes an online proactive caching approach to predict time-series content requests and update edge caching, where the future UE requests are predicted using convolutional recurrent neural networks. 
\cite{sadeghi2017optimal} further models UE preference over time at both local and global scales by Markov chain. Q-learning-based linear approximation function is designed to derive the optimum caching strategy in an online way, which can scale up to large wireless networks. Building on these prior researches, we introduce a novel approach termed DT-assisted data-driven optimization, utilizing a constrained Markov decision process (CMDP), which bridges the gap between stochastic optimization, lacking robustness to distribution errors, and robust optimization, which overlooks available problem data.  Notably, the utilization of created DTs enables the identification of specific moments through rich pre-generated data distributions and density functions. This approach also addresses the challenges associated with conventional data-driven methods, especially in scenarios where the precise distribution of random data variables remains unknown.

In the field of wireless caching, extensive data-driven studies have been made to improve the cache hit rate, save energy, and reduce transmission latency. However, the \textit{reliability} of edge caching, one of the stringent requirements of nextG networks, is rarely explored. Cache replacement decisions must prioritize system reliability by ensuring that BSs are balanced and not overloaded. For instance, if a specific BS is operating near its capacity limit, it becomes crucial to distribute or replace the corresponding content among other BSs that have sufficient capacity. This action ensures a balanced and optimized network, preserving the system's efficiency and stability. Failure to do so may result in severe consequences, including system crashes (e.g., resource exhaustion and network congestion) that jeopardize the overall network operations and significantly degrade the user experience.
Therefore, for the sake of the sustainability of network operations, it is imperative to integrate robust intervention mechanisms into data-driven optimization, being able to continuously monitor environment conditions, dynamically adjust caching strategies, and distribute the load efficiently among BSs. 

Motivated by the above limitations of traditional caching methods and conventional optimization approaches, as well as the insufficient attention given to reliability concerns within caching network systems,
we make the first attempt at reliable edge caching optimization, by combining reliable reinforcement learning (RL) with DTs. Multiple effective intervention modules are embedded into the data-driven optimization model to avoid network instability. Reliable RL algorithms facilitate trustful decision-making by accounting for reliability constraints and minimizing risks, which are learned from current input data and interactions with the environment. 
In parallel, DTs~\cite{9899718, ahmadi2021networked, zhang2024mapping}, the virtual replicas of physical networks, can be created and employed as RL optimizers and safeguards within the caching process. By imitating network behavior and accurately predicting the impact of caching decisions, DTs enable the optimization of caching strategies within a controlled environment. Furthermore, they act as \textit{reliability nets} by actively monitoring and verifying the integrity of cached content, ensuring immunity to malware and other potential threats. For example, DT can be employed to create \emph{rare} network scenarios, anticipating potential future challenges and aiding the network's transition from a \textit{data-driven} to a \textit{knowledge-driven} paradigm. 
\begin{figure}
	\centering
	\includegraphics[scale = 0.35]{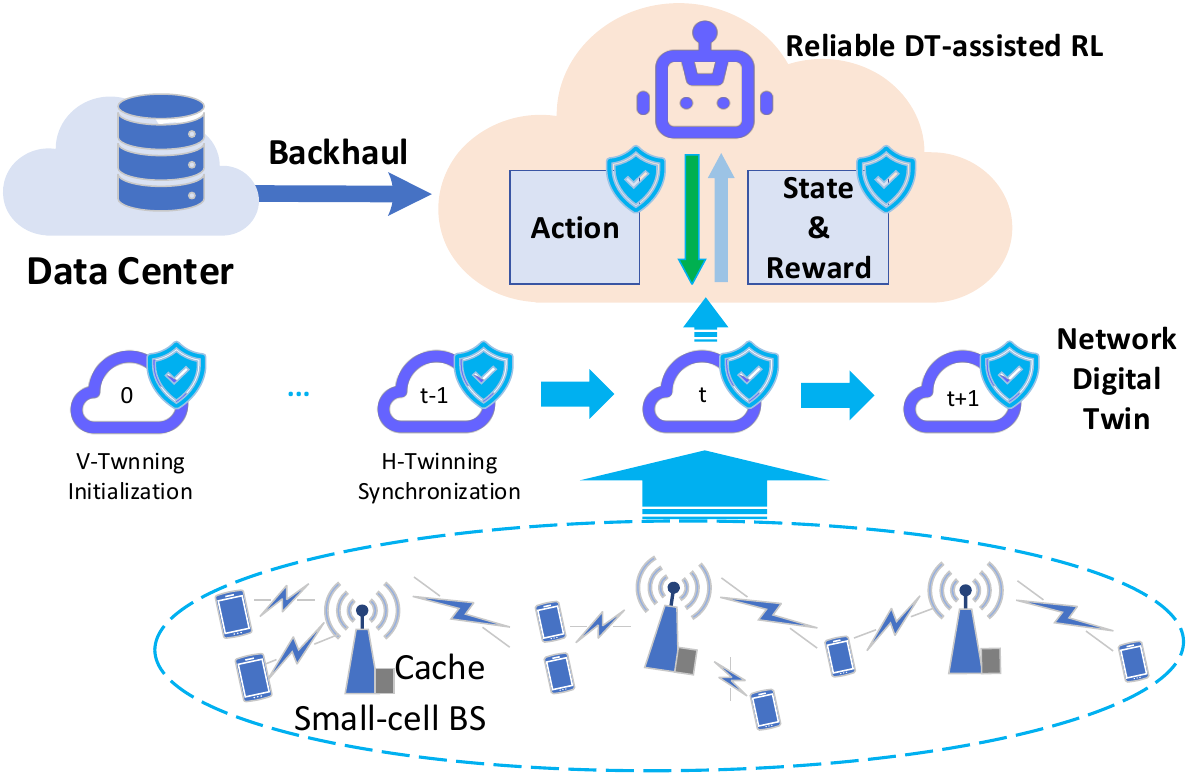}
	\caption{Overall caching optimization framework in cellular networks.}
	\label{fig: structure}
		\vspace{-.1in}
\end{figure}

Specifically, the contributions of this work are fourfold:
\begin{list}{\labelitemi}{\leftmargin=1em \itemindent=-0.0em \itemsep=.1em}

\item In this work, we introduce D-REC, the DT-assisted reliable RL mechanism for wireless caching optimization. Unlike existing approaches, our algorithm emphasizes incorporating on-demand constraints, including state, reward, and action safety modules, to prioritize network reliability and sustainability. This is the first work to investigate reliable data-driven method for caching optimization in nextG wireless networks.

\item We pioneer the creation and use of network DTs as data-driven optimizers and safeguards for wireless caching tasks. The diverse data distributions generated by DTs make the network aware of potential risks, allowing proactive countermeasures using our D-REC framework. This novel application highlights the versatility and potential of DT techniques in networking areas.

\item We present rigorous theoretical evidence demonstrating that the integration of reliability intervention modules into the caching optimization process does not affect the convergence performance of data-driven models. This confirms that our approach sustains optimal wireless caching performance while prioritizing network reliability.

\item The efficacy of our D-REC framework is validated through experiments on various modules and datasets. Results show that D-REC significantly improves the balance between BS load and resources without degrading cache hit rate, even under diverse user request distributions. These findings confirm the effectiveness of our data-driven optimization in practical network scenarios.

\end{list}

% !TEX root = mainfile.tex

\begin{table}[hbt!]
\centering
\caption{Related Notations and Definitions.}
\label{tab:notation}
\renewcommand{\arraystretch}{1.3} % Increases the row height
\begin{tabular}{|c|l|}
\hline
\multicolumn{1}{|c|}{\textbf{\textit{Notation}}} & \multicolumn{1}{c|}{\textbf{\textit{Description}}} \\ \hline
\( A \) & Action space in CMDP \\ \hline
\( a^t \) & Cache decision at time slot \( t \) \\ \hline
\( b \) & Base Station (BS) Load \\ \hline
\( B \) & Unreliable action intervention threshold \\ \hline
\( cap \) & Cache capacity of each BS \\ \hline
\( d \) & Specific content requested \\ \hline
\( H_n^t \) & Cache status of BS \( n \) at time slot \( t \) \\ \hline
\( h_{n,i}^t \) & Content stored in the \( i \)-th slot of BS \( n \) at time \( t \) \\ \hline
\( M \) & Frequency measure of requested content \\ \hline
\( N \) & Number of BSs \\ \hline
\( n \) & Index of current BS \\ \hline
\( P \) & Probability of random variable under particular distribution \\ \hline
\( p \) & Shape parameter of Zipf distribution \\ \hline
\( \pi_t \) & Policy at time \( t \) in Markov Decision Process \\ \hline
\( Q^{\pi} \) & Action-value function corresponding to policy \( \pi_K \) \\ \hline
\( r \) & Instantaneous reward function in CMDP \\ \hline
\( S \) & State space in CMDP \\ \hline
\( T \) & State transition probability matrix in CMDP \\ \hline
\( w \) & Instantaneous penalty function in CMDP \\ \hline
\( X \) & Random variable representing the rank of content \\ \hline
\end{tabular}
\end{table}

\section{Preliminaries and Related Work} \label{sec:related}

In this section, we provide an overview of the considered problem, relevant technologies, and related works, which includes the following topics: 
i) cache replacement problem, 
ii) relevant research on edge caching optimization, 
iii) reliable reinforcement Learning for caching,
and iv) digital twins and data generation. 
Important notations used in the paper can be found in Table~\ref{tab:notation}.

% \vspace{-0.1cm}
\subsection{Cache Replacement and Problem Formulation}
\begin{figure}
	\centering
	\includegraphics[scale = 0.3]{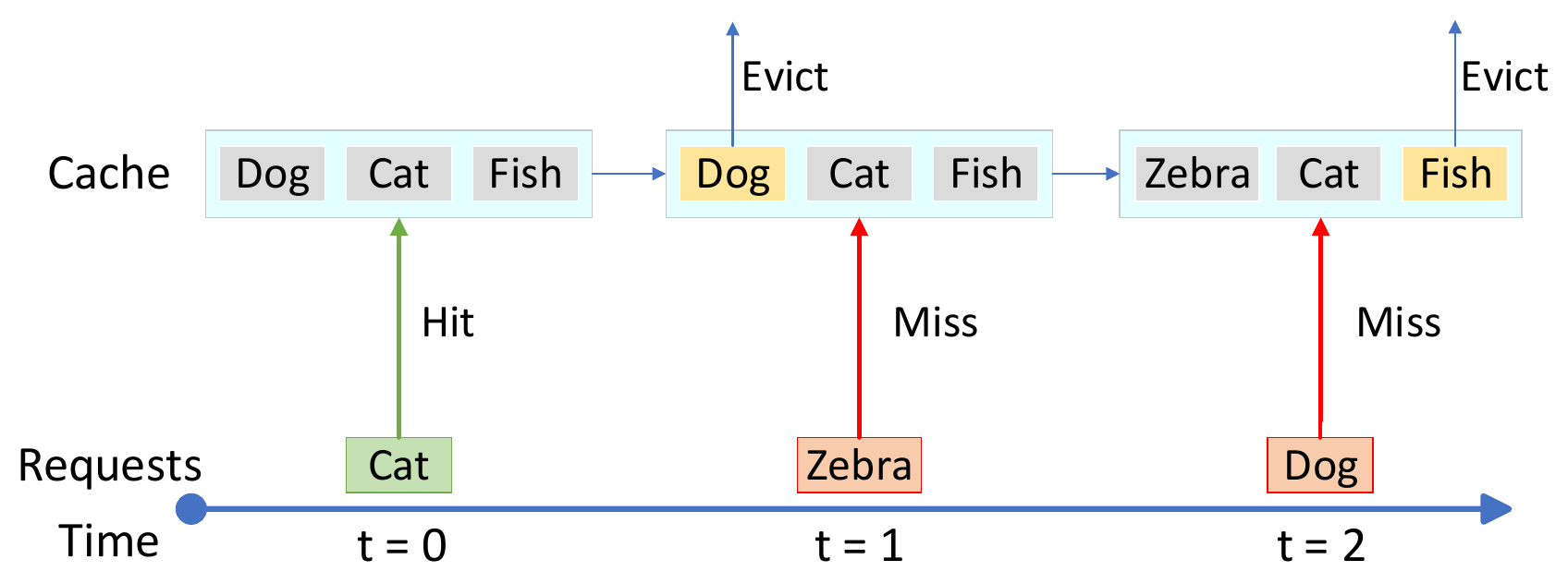}
	\caption{Example of a cache replacement problem.}
	\label{fig: cache}
		\vspace{-.1in}
\end{figure}

Emerging nextG cellular networks are expected to employ dense deployments of small-cell BSs. Consider a network composed of $N$ BSs, each interconnected with the same data center via low-bandwidth, high-delay backhaul links. Every BS is equipped with a cache unit of capability $cap$, capable of storing content retrieved from the data center. We introduce a symmetric cache scheme that assumes that all caching slots are of equal size and that each cache unit occupies the same amount of space. For the sake of simplicity, the contents are of cache capability, and in every time slot $t$, a certain number of contents will be requested by end users. As illustrated in Fig. \ref{fig: cache}, if the requested content is already cached in the BS cache unit (a situation referred to as a \emph{cache hit}), the content can be immediately downloaded. If not, the content must be fetched from the data center via backhaul links, inevitably leading to a higher communication cost and an increased transmission delay. The concept of cache replacement involves preemptively fetching certain contents via backhauling and storing them in the BSs before they are requested by users. Particularly, the optimization objective of this problem is twofold: 1) maximize the average cache hit rate and 2) minimize the peak traffic load on backhaul links.
In this regard, we address the problem of wireless edge caching as a joint optimization problem, aiming to maximize the cache hit rate while minimizing the overall delay and costs within the wireless network. 

Specifically, each BS \( n \) is equipped with a cache unit possessing a capacity of \( cap \) slots, designed to store contents for faster accessibility. A central server in the backend
processes requests from local BSs. A request refers to a read or write operation made to a cache unit in network settings. We define the cache status \( H_n^t \) of BS \( n \) at time slot \( t \) as follows:
\begin{equation}
    H_n^t = \{h_{n,1}^t, h_{n,2}^t, \ldots, h_{n,cap}^t\},
\end{equation}
where \( h_{n,i}^t \) represents the content \( d_i \) stored in the \( i \)-th slot of the cache unit at BS \( n \) at time \( t \). Initially, all cache slots are empty at \( t = 0 \). In the operational phase, if BS \( n \) has vacant slots in its local cache unit, it automatically accepts new requests and stores the corresponding content \( d^t \) in the empty slot. When the cache slots of BSs reach full capacity, the central server first decides whether to accept incoming requests. If a request is approved, the server then determines which specific slot should accommodate the new content \( d^t \).
The cache decision \( a^t \) is formulated as:
\begin{equation}
    a^t = \{ 0, 1, 2, \ldots, c \times n \}.
\end{equation}
A decision \( a^t = 0 \) implies retaining the current state of all cache slots unchanged. For \( a^t \neq 0 \), the designated cache slot will be cleared to make room for new content. The strategic objective for optimal cache decisions involves displacing less frequently accessed contents \(d\) from the cache units to enhance the overall hit rate and reduce latency in the network.

In line with previous research, the frequency of requested contents occurrence, denoted as $M$, typically follows a Zipf distribution, also known as a Zeta distribution~\cite{749260}. Formally, the Zipf distribution for a random variable \( X \) is represented as:

\begin{equation}
    P(X = k) = \frac{1/k^p}{\sum_{n=1}^{N_c} (1/n^p) },
\end{equation}

\noindent where \( k \geq 1 \) represents the rank of the frequency of requested content occurrence, and \( P(X = k) \) is the probability that the random variable \( X \) assumes the rank \( k \). The exponent \( p \), which characterizes the shape of the distribution, is always greater than zero. \( N_c \) denotes the total number of contents in the historical record. The term \( 1/k^p \) in Eq.~(3) signifies a power-law function, indicating that the probability is inversely proportional to the rank raised to the power of \( p \). The denominator, \( \sum_{n=1}^{N_c} (1/n^s) \), acts as a normalization factor to ensure that the sum of the probabilities across all ranks from 1 to \( N_c \). While the Zipf distribution is a common model in practical applications, other less frequent scenarios do exist. To enable data-driven optimization with knowledge of the diverse frequency of content occurrence distributions of data variables, DTs become a remedy for simulating those rarer cases, thereby enhancing the generality of the optimization model.

\subsection{Related Work on Edge Caching}
Edge caching strategies have transitioned from conventional methods to sophisticated machine learning-based approaches. Traditional caching mechanisms, such as Random Caching \cite{7011389}, Least Recently Used (LRU) \cite{10.1145/170035.170081}, Least Frequently Used (LFU) \cite{970573}, and Most Frequently Used (MFU) \cite{10.5555/1286760.1286772}, serve as foundational methods for addressing edge caching challenges. These strategies employ various selection criteria for caching or replacing content, ranging from random selection to prioritization based on access frequency or recency. Despite their simplicity and widespread adoption, these methods may not effectively adapt to the dynamic nature of network conditions and user demands.

Recent advancements in edge caching have increasingly utilized deep reinforcement learning (DRL) to optimize caching decisions in wireless networks. A notable study \cite{8964499} introduces a DRL-based edge caching scheme that dynamically adjusts cache content in response to network conditions and user demands, leading to enhanced cache hit rates and reduced latency. Wang et al. \cite{10.1145/3623398} propose an intelligent cooperative caching strategy at the mobile edge, leveraging offline DRL to improve the performance of cooperative caching in mobile edge networks. Additionally, Basic DQN \cite{wang_drlcache_2023} employs a DQN model for cache replacement decisions, showcasing the potential of advanced DRL approaches in edge caching.
Beyond DRL-based methods, collaborative and content-based caching strategies have also been explored. \cite{9166756} presents an online collaborative data caching framework for edge computing, enabling edge nodes to share cached content and thereby enhance system performance. In \cite{WEI2021102000}, a wireless edge caching scheme is proposed to capitalize on content similarity for boosting cache hit rates in dynamic environments. Moreover, the integration of federated learning in edge caching, as explored in \cite{10015857}, offers a promising avenue for optimizing caching while preserving user privacy through distributed learning techniques. 
The exploration of multi-agent reinforcement learning (MARL) for edge caching represents another significant direction. Research in \cite{10123387} employs MARL to facilitate cooperative caching in the Internet of Vehicles, demonstrating the efficacy of distributed learning approaches in complex network scenarios. Collectively, these studies highlight the critical role of intelligent caching mechanisms, whether through collaboration, content analysis, or distributed learning, in improving the performance of edge networks.
By contrast, our work introduces a versatile framework designed to seamlessly integrate with a broad range of existing DRL models. As a foundational model, we employ basic DQN to construct a more advanced, data-driven, and reliable optimization framework specifically tailored for edge caching.

% \vspace{-0.4cm}

\subsection{Reliable Reinforcement Learning as a Primer}

Numerous cache replacement policies have been proposed over decades of research. Three of the most well-known heuristic approaches are Least Recently Used (LRU), Least Frequently Used (LFU), and Most Recently Used (MRU), which operate based on the principles of recency and frequency, respectively~\cite{zahran2007cache}. These traditional methodologies can be categorized as \emph{reactive} caching approaches. However, they neither account for the patterns of content frequency nor the cooperative interactions among BSs, leading to potential inefficiencies. In contrast, proactive caching strategies have gained increasing popularity in recent years, which first estimate content request patterns, and then determine the optimal policy accordingly. Typically, the cache replacement problem can be formulated as a Markov decision process and resolved using deep RL as we consider herein. Deep RL is an effective data-driven optimization method for complex nonlinear mapping relationships between states, actions, and other constraints. Such intelligent approaches can accommodate more complex and varying input data and scenarios, including considerations of content frequency in both temporal and spatial domains~\cite{sadeghi2017optimal}, cooperation between BSs~\cite{rezaei2018multi,wu2021multi}, and the particularities of heterogeneous mobile edge networks~\cite{nomikos2022survey}. However, the deployment of RL models in real-world applications often raises reliability concerns. Without timely intervention and appropriate safeguards, RL models may exhibit unexpected behaviors in certain circumstances, e.g., overloading a BS or repeating caching the same contents. Such irregularities possess the potential to undermine the efficacy of the caching policy. In more severe scenarios, they could even strain the network's traffic handling capabilities and consequently jeopardize the overall network stability and operation.

To address this concern, safety-oriented RL approaches~\cite{garcia2015comprehensive,brunke2021safe} have been developed to ensure reliability and stability throughout both the learning and deployment stages of data-driven systems.
Safe RL aims to develop intelligent agents that can not only maximize their performance over time but also operate within specific safety constraints, minimizing harmful actions and mitigating risks~\cite{alshiekh2017safe,turchetta2021safe}. These constraints may arise from a variety of contexts, including avoiding damage to the learning agent itself, preventing harm to other entities in the environment, or adhering to predefined operational guidelines. The ultimate goal of safe RL is to create a balance between exploration (learning about the environment to improve future performance) and exploitation (using current data and knowledge to maximize immediate performance), while keeping the system within safe operational boundaries, which can be well applied to various high-stake areas such as robotic systems and autonomous vehicles~\cite{brunke2021safe, Du_Ye_Gu_Li_Wei_Wang_2023, Singh2021BuildingHR}. Motivated by these previous findings, this work aims to tackle the challenge of designing and implementing reliability intervention models into our wireless caching optimization problem. These models should be capable of understanding and adhering to specific reliability boundaries, even when exploring unknown and edge cases.

\subsection{{Digital Twin as Data Generator for What-if Analysis}}

While safe RL offers a pathway to addressing some of the security concerns associated with edge caching, its efficacy is heavily contingent upon the availability of extensive data that accurately reflects the dynamic state of real-world wireless networks. Collecting such data poses significant challenges, not only due to the associated costs but also because of the practical difficulties in capturing a comprehensive representation of network conditions. Furthermore, validating the reliability of RL models before their deployment requires the simulation of a wide array of scenarios, a task that is both resource-intensive and operation-complex. In this context, DTs emerge as a highly promising solution to resolve these challenges.

Specifically, DTs can be conceptualized as virtual replicas of physical systems within the real world, capturing their key features and dynamic characteristics with two-way communications. These digital counterparts facilitate a continuous flow of data to and from their physical analogs, enabling real-time decision-making and improvements~\cite{khan2021digitaltwinenabled, li2024JSTSP, zeb2022industrial, Yang2024ICASSP}. The effectiveness of DTs hinges on seamless communication and precise modeling. In pursuit of these objectives, cutting-edge techniques such as machine learning and multi-physics simulations have been employed in DT development, which enable real-time status monitoring while simulating potential future states and delivering predictive insights~\cite{zhang2024mapping, lin20236g, liu2022environment}.

In wireless networks, the emergence of network DTs holds significant potential for the rapid development of 6G and beyond networks, particularly within the framework of Industry 4.0. With the expected proliferation of devices in nextG networks, there is a growing need to establish a scalable, reliable architecture rooted in DT technology. For instance, in~\cite{9491087}, the authors propose a wireless edge model that marries DT technology with edge networks. This integration yields novel functionalities such as hyper-connectivity and low-latency edge computing. Furthermore, in ~\cite{9174795}, a mobile offloading scheme is introduced for DT-enabled networks to reduce offloading latency, adhering to constraints surrounding the accumulated service migration costs incurred during user mobility. Blockchain, in conjunction with DTs, could enhance the security measures within 6G networks as outlined in ~\cite{9076112}. In another significant contribution from~\cite{8822494}, the authors introduce a DT designed for industrial automation and control systems. They also articulate the security requirements for data sharing and control based on DTs, underlining the need for robust data-driven security measures.
However, current studies on DTs tend to focus more on their applications rather than the processes involved in creating and synchronizing DTs with physical networks. In this work, our emphasis is on a complete cycle of employing DTs in wireless network optimization, encompassing creation, synchronization (Sec. III), applications (Sec. V), and validation (Sec. VII). 

\section{Creation and Synchronization of Network Digital Twins} 
\label{sec:dt}

Prior to utilizing DTs for our data-driven optimization, this section introduces a novel framework for creating and synchronizing network DTs, extended from~\cite{zhang2024mapping}. The overall approach is structured around three main stages: dynamic connectivity segmentation (DCS), vertical twinning (V-Twinning), and horizontal twinning (H-Twinning). Specifically,
DCS is employed periodically to ensure effective clustering on densely deployed BSs, V-Twinning is performed at the beginning to initialize a concrete global network digital twin (G-NDT), and H-Twinning is later performed to update the twins regularly.

\subsection{Dynamic Connectivity Segmentation}

\begin{algorithm}[t]
\caption{Dynamic Connectivity Segmentation (DCS)}
\begin{algorithmic}[1]
\Require relationship matrix \( \{ g, k, \beta, \tau \} \), attribute weights \( \omega \), number of clusters \( C \), number of BSs \( N \)
\Ensure Clusters \( c \)
\State Initialize \( \Phi\);
\For{\( n_1 = 1, 2, \ldots, N \)}
    \For{\( n_2 = 1, 2, \ldots, N \)}
      \State \( \Phi_{n_1,n_2} \gets \frac{\omega_g}{g_{n_1,n_2}} + \omega_k \cdot k_{n_1,n_2} + \omega_\beta \cdot \beta_{n_1,n_2}\)
      \State ~~~~~~~~~~~~\(+ \omega_\tau \cdot \tau_{n_1,n_2}\)
    \EndFor
\While{desired \( C \) clusters not reached}
    \State Calculate betweenness centrality for all edges;
    \State Remove edge with highest betweenness centrality;
    \State Recalculate the communities;
\EndWhile
\EndFor
\State \Return \( c \)
\end{algorithmic}
\label{algo:DCS}
\end{algorithm}

The DCS algorithm, as outlined in Algorithm~\ref{algo:DCS}, is designed to cluster multiple BSs responsible for network service areas with similar communication characteristics and networking configurations. This clustering step is integral to the efficient creation and updates of multiple distributed network DTs, i.e. cluster network digital twin (C-NDT), which demonstrate distinct behaviors and perform paralleled synchronization with the G-NDT.
This algorithm is executed periodically, ensuring dynamic clustering and thereby enhancing the twinning performance in real-time. 
In the DCS algorithm, clusters are based on an attribute sequence of BSs \( \{ g, k, \beta, \tau \} \), which represents their geological distances, capacity of backhaul links, coverage area overlaps, and similarity of frequency of occurrence distribution, respectively. 
Each BS and its corresponding NDT have their designated coverage area to provide services to clients. Overlapping coverage areas between BSs and NDTs indicate potential similarities in the services and requests from clients within these network regions. Additionally, similarity in the frequency of occurrence distribution suggests that client requests serviced by both BSs and NDTs are comparable. Typically, when both BSs and NDTs handle similar client requests, it increases the efficiency of wireless network management. This involves optimizing edge caching strategies to enhance network efficiency, reduce latency, and ultimately improve the overall user experience.
Initially, a relationship matrix incorporating these attributes, weighted by \( \omega \), is constructed. 
For BS \( n_1 \) and BS \( n_2 \), the algorithm calculates a metric \( \Phi_{n_1,n_2} \) to quantify their correlations using the formula as:
\begin{equation}
    \Phi_{n_1,n_2} = \frac{\omega_g}{g_{n_1,n_2}} + \omega_k \cdot k_{n_1,n_2} + \omega_\beta \cdot \beta_{n_1,n_2} + \omega_\tau \cdot \tau_{n_1,n_2},
\end{equation}
where $\omega$ is the weight to balance the significance of each attribute. Betweenness centrality is then computed based on \( \Phi \), which measures how often an edge, i.e. connectivity relationship, acts as a bridge along the shortest paths in the network. Edges with high betweenness centrality typically connect different communities. The algorithm clusters BSs by iteratively removing edges with the highest betweenness centrality until the desired number of clusters \( C \) is reached, which is similar to the Girvan-Newman method~\cite{Newman_2004}. 
This clustering process is crucial in determining how twinning models from local BSs are shared with their corresponding C-NDTs, ultimately contributing to the update of G-NDT, as discussed in the subsequent subsections. 
In general, regularly clustering in dynamic environments offers significant advantages, including reduced communication overhead. By grouping similar devices or clients, the frequency of communication with the central server is minimized, leading to more efficient network usage. In this way, enhanced model performance can be achieved as local models, trained on data from each cluster, are better representative of the specific data distribution within that cluster. 
Besides, resource utilization is naturally optimized by allocating computational resources to groups of clients with similar data distributions, which accelerates convergence and lowers operational costs. 
Furthermore, such clustering process enhances the robustness of non-IID data by grouping clients with similar data distributions, resulting in more stable and reliable twin models.

\subsection{Vertical Twinning for Initialization}

\begin{algorithm}[t]
\caption{Vertical Twinning (V-Twinning)}
\begin{algorithmic}[1]
\Require Local models \( \bm{\alpha}_1^t, \bm{\alpha}_2^t, \ldots, \bm{\alpha}_{N}^t \), number of clusters \( C \)
\Ensure Updated G-NDT \( \bm{\alpha}^{t+1} \)
\For{\( c = 1, 2, \ldots, C \)}
    \State \( P \gets \text{BS in cluster } c \)
    \State \(\bm{\alpha}_c^{t} \gets \frac{1}{P}\sum_{p=1}^P \bm{\alpha}_p^{t} \)
\EndFor
\State \( \bm{\alpha}^{t+1} \gets \frac{1}{C}\sum_{c=1}^C \bm{\alpha}_c^{t}\)
\State \Return \( \bm{\alpha}^{t+1} \)
\end{algorithmic}
\label{algo:VT}
\end{algorithm}

The V-Twinning stage aims to create initial network DTs with historical data on caching requests and their frequency. It employs a federated learning (FL) strategy, specifically tailored for wireless networks with multiple BSs organized in clusters. FL is a machine learning technique where model parameters are shared among BSs instead of raw data, enabling collaborative training of a global model. This approach efficiently distributes twinning tasks across BSs while ensuring content data privacy. As depicted in Algorithm~\ref{algo:VT}, historical caching data from each BS \( n \) are used to train C-NDTs for each cluster \( c \) first. With local twin models shared from BSs within the same cluster, denoted as \(\bm{\alpha}_1^t, \bm{\alpha}_2^t, \ldots, \bm{\alpha}_N^t\), the corresponding C-NDT \(\bm{\alpha}_c^t\) aggregates the models to reach a consensus. The most common aggregation rule FedAvg~\cite{mcmahan2017communication} can be used to compute the dimension-wise arithmetic mean of each twinning model parameter.

G-NDT, represented as \(\bm{\alpha}^{t+1}\), is the averaged aggregator of multiple C-NDTs at $C$ clusters, i.e.,  \(\bm{\alpha}^{t+1} = \frac{1}{C}\sum_{c=1}^C \bm{\alpha}_c^t\), where \(C\) is the number of clusters. Then, the model parameters of G-NDT are sent back to each cluster for synchronizing C-NDTs after the twinning aggregation process.

Specifically, V-Twinning employs synchronous FL, ensuring that all DTs update their twinning models simultaneously. This mechanism is crucial for maintaining a consistent and stable twinning process across the network areas. 
By synchronizing the model updates from all participating clients at regular intervals, a collaborative learning process is guaranteed, leading to potentially more stable and predictable performance. Additionally, the use of synchronous FL for twinning can simplify the management of model updates and reduce issues related to stale or incompatible data, making it suitable for scenarios where uniformity and coordination among BSs are critical.

\subsection{Horizontal Twinning for Evolution}

\begin{algorithm}[t]
\caption{Horizontal Twinning (H-Twinning)}
\begin{algorithmic}[1]
\Require Local models \( \bm{\alpha}_1^t, \bm{\alpha}_2^t, \ldots, \bm{\alpha}_{N}^t \), current G-NDT \( \bm{\alpha}^{t} \), number of clusters \( C \), threshold \( \psi \)
\Ensure Updated G-NDT \( \bm{\alpha}^{t+1} \)
\For{\( c = 1, 2, \ldots, C \textbf{ asynchronously}\)}
    \State \( P \gets \text{BS in cluster } c \)
    \State \(\bm{\alpha}_c^{t} \gets \frac{1}{P}\sum_{p=1}^P \bm{\alpha}_p^{t} \)
    \State \( \epsilon \gets (\bm{\alpha}_c^{t} + \bm{\alpha}^{t})^2 \)
    \If{\( \epsilon > \psi \)}
        \State \( \bm{\alpha}^{t+1} \gets \frac{1}{C}\sum_{c=1}^C \bm{\alpha}_c^{t}\)
    \Else
        \State \( \bm{\alpha}^{t+1} \gets \bm{\alpha}^{t}\)
    \EndIf
\EndFor
\State \Return \( \bm{\alpha}^{t+1} \)
\end{algorithmic}
\label{algo:HT}
\end{algorithm}

To ensure the network DTs remain relevant, H-Twinning stage is designed to periodically synchronize between the physical twin and DT with real-time data. Unlike V-Twining, it adopts an asynchronous FL approach to update with dynamics from the physical twin, aiming to provide a scalable and flexible solution for wireless networks composed of multiple clusters.

As described in Algorithm \ref{algo:HT}, H-Twinning begins with $N$ local models \(\bm{\alpha}_1^t, \bm{\alpha}_2^t, \ldots, \bm{\alpha}_N^t\) from respective BSs and the current G-NDT \(\bm{\alpha}^t\). The use of the threshold \(\psi\) serves as a criterion to decide whether the G-NDT should be updated. It assesses the deviation between a C-NDT \(\bm{\alpha}_c^t\) and the current G-NDT \(\bm{\alpha}^t\), quantified by \(\epsilon = (\bm{\alpha}_c^t - \bm{\alpha}^t)^2\). If \(\epsilon\) surpasses the threshold \(\psi\), indicating a significant change in the physical network, the G-NDT is updated to reflect the fresh information. The updated G-NDT, \(\bm{\alpha}^{t+1}\), is calculated as an average of C-NDTs at the current time slot, \(\bm{\alpha}^{t+1} = \frac{1}{C}\sum_{c=1}^C \bm{\alpha}_c^t\). If \(\epsilon\) is within the threshold \(\psi\), the G-NDT remains with the current model, i.e. \(\bm{\alpha}^{t+1} = \bm{\alpha}^t\). This threshold-based update mechanism enhances the network's efficiency by ensuring that only significant changes will lead to twin updates, thereby reducing unnecessary computational overhead and preserving bandwidth.

Compared with V-Twinning, H-Twinning does not require simultaneous updates from all clusters, although the overall procedure to compute C-NDT and G-NDT is quite similar. Each BS \( n \) stores the real-time data stream first and trains the twin model when the amount of data reaches a threshold in batches. Asynchronous FL naturally offers several advantages over its synchronous counterpart in such a twinning problem. Primarily, it allows for greater flexibility in participation, as BSs and network DTs can contribute to the model training process at their own pace and availability, without being bound to a strict synchronization schedule. This feature is particularly beneficial in wireless scenarios with BSs having varying computational resources or backhaul connectivity, ensuring that the twinning process is inclusive and efficient even in less ideal conditions.

With the above three-stage twinning process, we implement a complete case study to create network traffic twins for caching optimization using real-world wireless data, as detailed in~\cite{dataset}. This case study demonstrates the capability of a network DT to accurately simulate and predict wireless traffic patterns. This predictive capability paves the way for anomaly detection and system protection purposes. Due to space constraints, we omit details here, but interested readers can refer to our technical repository at~\cite{dnt}.

In summary, the overall distributed NDT system can effectively address the challenges of heterogeneity, randomness, and variability of user devices in wireless networks. By employing the DCS technique, BSs and NDTs sharing similar characteristics and data patterns are grouped together, thereby reducing the heterogeneity within each C-NDT. This facilitates the creation of more homogeneous and representative local twin models.
Following this segmentation, the appropriate aggregation techniques are adopted to achieve a consensus across these clusters and local twin models. This aggregation process can further reduce heterogeneity within the network. Specifically, by combining the local models of each cluster into a global twin model, which is G-NDT, we can address the disparities between different clusters and local twin models. This not only harmonizes the network's overall behavior but also enhances the efficiency and accuracy of the mapping process. By mitigating such heterogeneity, the aggregation in the proposed FL framework ensures that the global model is more representative and robust, leading to enhanced performance and reliability in the network operations.

% !TEX root = mainfile.tex

\section{Caching Optimization with Constrained Markov Decision Process} 
\label{sec:system_design}

In this section, we formulate the data-driven optimization problem of cache replacement using a constrained Markov decision process (CMDP) and then present an RL-based solution.

\subsection{Problem Formulation with CMDP Model\label{sec:rl_model}}

A constrained Markov decision process (CMDP) extends the traditional MDP framework, commonly utilized in modeling decision-making within stochastic environments. Diverging from standard MDPs that primarily aim to maximize cumulative reward, CMDPs integrate additional constraints into the decision-making framework. These constraints, typically in the form of limits on specific metrics or resource usage, ensure solutions not only optimize the primary objective but also comply with predefined bounds. CMDPs are thus ideal for scenarios necessitating a balance between multiple objectives or adherence to operational constraints~\cite{gu2023review}, as in our wireless caching context. The process is characterized by states, actions, transition probabilities, a reward function, and constraints on expected cumulative costs or rewards. 
Particularly, the objective of a CMDP is to maximize long-term cumulative reward while adhering to these constraints.
In this regard, we formulate our caching optimization problem, as introduced in Sec.~\ref{sec:related}-A, into a CMDP model \( \{S, A, r, w, T, s^0\} \) comprising the following elements:
\begin{itemize}
    \item \( S \) is the state space, comprising a tuple \((n, t_n, f_n)\), where \( n \) denotes BS index. \( t_n \) and \( f_n \) represent the last time the content was cached and its frequency, respectively. This can be expressed as $S = \{(n, t_n, f_n) \mid n \in N, t_n \in t_{\text{total}}, f_n \in \mathbb{N}\}.$
    \item \( A \) is the action space, detailing the possible cache decisions as outlined in Sec.~\ref{sec:related}. The central server decides whether to accept a request and subsequently selects a cache slot to store the request \( d^t \). Mathematically, it can be represented as $A = \{ 0, 1, 2, \ldots, c \times n \},$
    where 0 means skipping this request, while other values represent the index of the cache slot for caching.
    \item \( r \) is the instantaneous reward function, quantifying the number of cache hits between two request acceptances, which are $hit(t)$ and $hit(t+1)$. It can be formulated as:
    \begin{equation}
        r(s, a) =  \upsilon \cdot (hit(t+1) - hit(t)),
    \end{equation}
    where $\upsilon$ is a hyperparameter to adjust the significance of the reward.
    \item \( w \) represents the instantaneous penalty, accounting for cache misses caused by prior actions, which can be defined as:
    \begin{equation}
        w(s, a) = \frac{1}{\left[r(s, a)\right]^\kappa + 1},
    \end{equation}
    where $\kappa$ is a hyperparameter to control the significance of the penalty.
    \item \( T \) denotes the state transition probability matrix, describing the probabilities of moving from one state to another based on specific actions.
    \item \(s^0 \in S\) signifies the initial state after all cache units are full, serving as the starting point of the CMDP sequence.
\end{itemize}

In this way, given caching policy \(\pi\), the long-term cumulative reward \(R\) and long-term cumulative cost \(W\) can be derived as follows:
\begin{equation}
    R_\pi(s^0) = \mathbb{E} [ \sum_{t=0}^{t_{\text{total}}} (r(s^t, a^t) | s^0, \pi ],
\end{equation}
\begin{equation}
    W_\pi(s^0) = \mathbb{E} [ \sum_{t=0}^{t_{\text{total}}} (w(s^t, a^t) | s^0, \pi ].
\end{equation}

\noindent Furthermore, to determine a set of reliable intervention policies that satisfy the specific constraints in CMDP, an approximated auxiliary cost parameter $\mu$ can be added to derive a Lyapunov function as follows:
\begin{equation}
    L_{\mu} = \mathbb{E} \left[ \sum_{t=0}^{t_{\text{total}}} w(s^t, a^t) + \mu \,\middle|\, s^0, \pi \right]
\end{equation}

\subsection{RL-based Solution}

Edge caching involves managing the storage and retrieval of data at the edge of a network, positioned closer to end-users, with the aim of reducing latency and alleviating network congestion. This requires the optimization of several factors, including cache placement, sizes, replacement policies, and content popularity, all of which can vary over time and across different network conditions.
RL excels in optimizing such complex mapping relationships due to its ability to learn and adapt from dynamic, non-linear interactions within an environment, effectively navigating and optimizing decisions based on trial and error without requiring predefined models. 
This makes RL such as Deep Q Networks (DQN) particularly adept at handling intricate, multi-dimensional optimization problems where traditional algorithmic approaches might struggle to capture the nuanced interdependencies and variations.
Our objective is to offer a comprehensive solution for edge caching challenges and to implement dependable modules that are compatible with various RL algorithms. We utilize DQN as our foundational model to address the formulated problem as in Sec. IV-A. Additionally, the developed optimization approach can be readily adapted to accommodate other RL algorithms, ensuring versatility and broad applicability in solving edge caching issues.

Specifically, the state-action reward function \( Q_r \) and the state-action cost function \( Q_w \) are pivotal in navigating this optimization landscape, mapping state-action pairs to expected rewards and costs, respectively. 
The reward function \( Q_r \), central to our approach, aims to quantify the expected cumulative reward from a specific state-action pair over the remaining time horizon. This is formalized as:
\begin{equation}
Q_r(s_m, a_m) = \mathbb{E}\left[\sum_{t=m}^{t_{\text{total}}} \gamma^{t-m}r(s_t)\middle|s^0, a^0\right],
\end{equation}
where \(\gamma \in [0,1]\) is the discount factor, moderating the value of future rewards and embodying the principle of time preference. 
This temporal weighting is crucial in our edge caching problem, where immediate access to frequently requested content is more valued, thus prioritizing cache decisions that cater to current demand patterns.

The expected reward and cost value functions, \( V_r^{\pi} \) and \( V_w^{\pi} \), respectively, are determined by both immediate returns and the discounted value of future states, guiding the policy towards optimal caching strategies. For instance:
\begin{equation}
Q_r(s, a) = r(s) + \gamma V_r^{\pi}(s'), \quad \forall s \in S, a \in A,
\end{equation}
highlighting how immediate rewards and anticipated future benefits inform the caching decisions to ensure content availability and optimal resource allocation.

Similarly, the cost function \( Q_w \) considers both immediate costs and future expenses under a policy \( \pi \), reflecting the cumulative cost impact of caching decisions:
\begin{equation}
Q_w(s, a) = w(s) + \gamma V_w^{\pi}(s'), \quad \forall s \in S, a \in A,
\end{equation}
This aspect is vital in managing resource constraints within the considered edge network, ensuring that caching strategies do not exceed storage capacities or degrade network performance.

The Lyapunov function \( L_{\mu} \), integral to maintaining system stability and adherence to operational constraints, further refines this optimization process by incorporating the cumulative cost and above constraints, making it particularly suitable for dynamic caching environments.

\begin{equation}
L_{\mu}(s, a) = w(s) + \mu + \gamma L_{\mu}^{\pi}(s'), \quad \forall s \in S, a \in A,
\end{equation}

The ultimate goal is to identify an optimal policy \( \pi^* \) that not only maximizes expected rewards but also complies with the network's constraints, crucial for sustaining edge caching efficacy and reliability:
\begin{equation}
\pi^*(\cdot|s) = \arg \max_{\pi(\cdot|s) \in F_{L_{\mu}}(s)} \pi(\cdot|s)^T Q_r(s, \cdot), \quad \forall s \in S,
\end{equation}

Through this refined approach, we can effectively address the heterogeneity, randomness, and variability of user devices in wireless networks by leveraging a data-driven, adaptive framework that optimizes edge caching decisions, enhancing both content delivery and network performance.

\section{D-REC: Reliability Module Integration}
\label{sec:modules}

\begin{figure*}[hbt!]
    \centering

    \subfloat[\footnotesize{State module}]{\label{subfig:a} \includegraphics[width=0.32\textwidth]{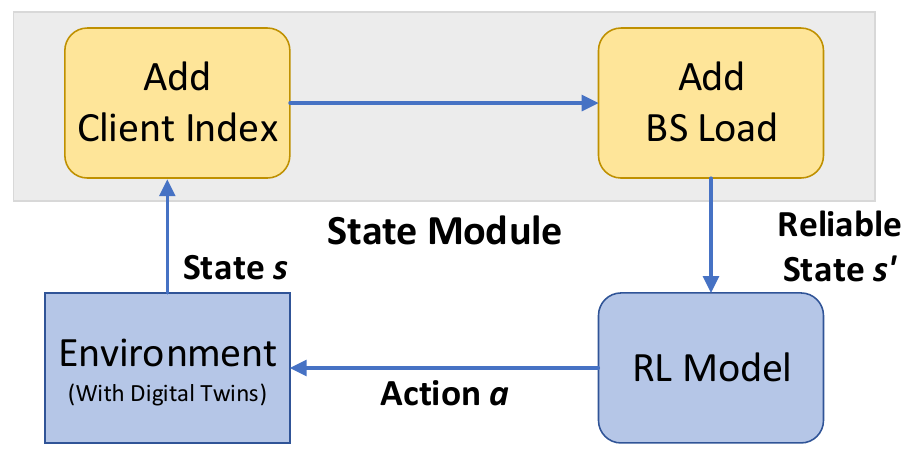}}
    % \hfill
    \subfloat[\footnotesize{Action module}]{\label{subfig:b} \includegraphics[width=0.32\textwidth]{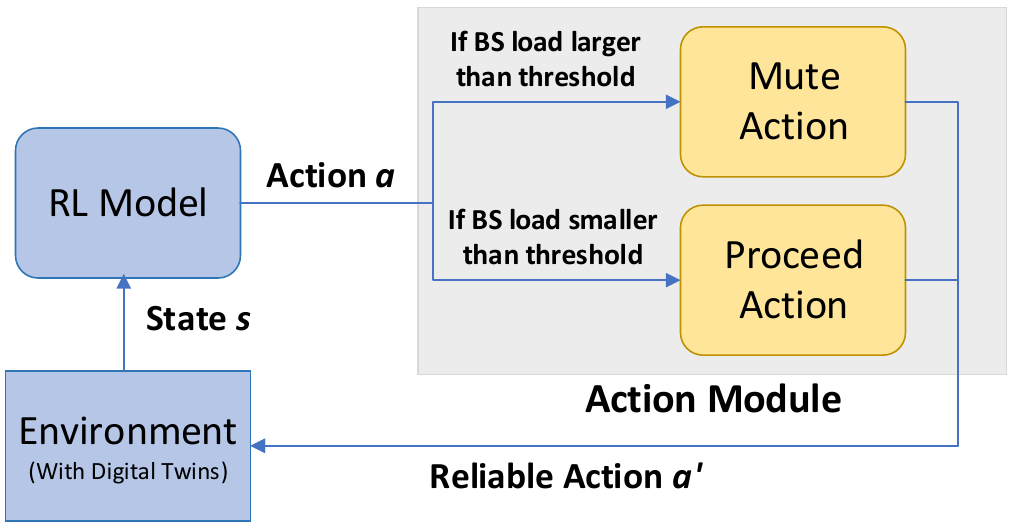}}
    % \hfill
    \subfloat[\footnotesize{Reward Module}]{\label{subfig:c} \raisebox{-.05\height}{\includegraphics[width=0.3\textwidth]{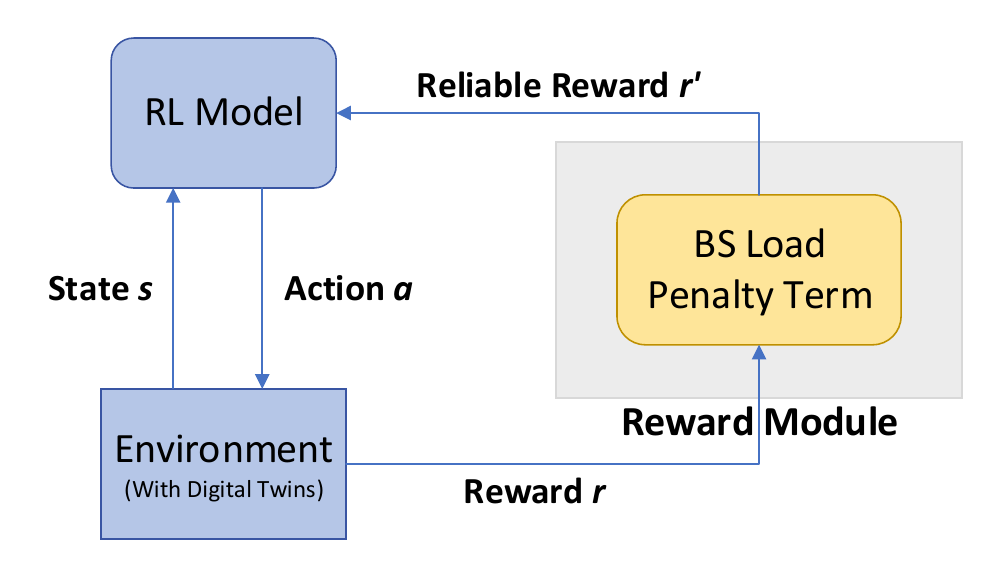}}}
    \caption{A zoom-in view of intervention modules. The grey box indicates the effective area of reliability intervention modules.}
    \label{fig: saferl}
    \vspace{-.1in}
\end{figure*}

Despite the promising capabilities of the RL model as a solution for our reliable edge caching (REC) problem, we have encountered persistent challenges concerning its reliability in practice network operations. Consequently, there is a substantial need for incorporating reliability interventions specifically designed to safeguard the efficiency, stability, and overall reliability of the entire network system. To tackle these issues, we propose four innovative reliability intervention modules integrated into our DT-assisted data-driven optimization. Each module is meticulously designed to bolster system reliability and efficiency without compromising achievable network performance. The design and operational details of these reliability modules are presented in Fig.~\ref{fig: saferl}.

\vspace{+0.1cm}
\subsection{REC with Network Digital Twin (D-REC)}

We employ the built network DT in our optimization framework as both an RL optimizer and a safeguard, where real-time synchronization can be ensured through two-way communications between the physical network and DTs through the proposed V-H twinning methods.
Specifically, the V-Twinning framework detailed in Sec.~\ref{sec:dt} involves supplying network DTs with historical caching data, including information about requests, contents, frequency, and other essential attributes. Then, through the utilization of data-driven techniques, such as long short-term memory, DTs are capable of producing one-step forecasts for the subsequently requested contents. Following content forecasts, we analyze the distribution of content occurrences. Various sets of historical caching data can be employed to feed DTs, generating datasets that cover both common and rare wireless scenarios. Leveraging the frequency distribution of content occurrences, we can then generate content based on their likelihood of occurrence.
These contents and their associated attributes serve as inputs to the RL model as detailed in Sec.~\ref{sec:rl_model}, acting as states within the algorithm. Upon the decision-making process of the RL model, real-time data is transmitted to network DTs for the execution of the H-Twinning stage, to maintain an accurate twinning model in a closed loop.
The DT-assisted approach is further enhanced by incorporating the following reliability intervention modules, specifically designed to proactively address potential network risks and ensure a robust optimization process.

\vspace{+0.1cm}
\subsection{D-REC with State Intervention Module}

In Sec.~\ref{sec:rl_model}, we define the state $s$ in the CMDP as $(n,~g_n,~f_n)$, which represents the BS index, the last time that the cache line was cached, and its caching frequency, respectively. Here, a reliability module is added to extend the state space of our RL-based solution. As shown in Fig.~\ref{fig: saferl}(a), we incorporate two additional variables into the cache state, namely, the client index $j$ that is currently requesting content, and the normalized load $\hat{l}$ for the $n$-th BS. The normalized load $d_n$ is the proportion of requests handled by the $n$-th BS out of all requests at time $t$. These additional variables force the policy model to consider the load of BSs in the wireless network. Therefore, the RL model can detect potential overloading issues of BSs, leading to more effective and balanced cache management.

\vspace{+0.1cm}
\subsection{D-REC with Reliable Action Module}

As illustrated in Fig.~\ref{fig: saferl}(b), the caching actions taken by the RL model can be secured with a manually designed \emph{backup} policy, i.e., dividing them into mute actions and proceed actions. When a target BS is overloaded, the action generated by the RL model will be \emph{muted} by chance. Formally, the $n$-th BS is regarded as overloaded if 
\begin{equation}
    \hat{b} - \min_{y} b_y \ge B,
\end{equation}
where the unreliable action intervention threshold $B$ (e.g., set as 0.2) is a hyperparameter and $b$ represents the current BS load. In particular, BS load quantifies the volume of requests handled by a given BS at any specific moment and is expressed in percentage. 
Such a flexible design ensures that the decisions of the RL model are not blindly enforced but allow for effective intervention when necessary.

\vspace{+0.1cm}
\subsection{D-REC with Reward Intervention Module}

Furthermore, an additional penalty term is integrated into the original reward function in Eq.~(10) (see Fig.~\ref{fig: saferl}(c)) to safeguard the caching process. After computing the reward from the RL model, we penalized the model by the average difference between $n$-th BS and the minimum BS load, which can be formulated as follows:
\begin{equation}
    r(s,a)' = r(s,a) - |\frac{\phi}{N} \sum_{n} (\hat{b} - \min_{y} b_y)|,
\end{equation}
where $N$ is the total number of BSs in the network, $\phi$ is a hyperparameter to balance the significance of this additional term, $s$ and $a$ are the corresponding state and action, respectively. Such a penalty term instructs the RL model to prioritize both balance in BS load and caching efficiency throughout the optimization process.

Overall, all the aforementioned intervention modules have different positive impacts on the data-driven optimization process, which will be analyzed and evaluated in the subsequent sections.

% !TEX root = mainfile.tex

\section{Theoretical Analysis of D-REC Optimization} 
\label{sec:theoretical_results}

This section provides a theoretical analysis of the proposed reliability intervention modules. We demonstrate that these integrated modules have no impact on the convergence rate of the data-driven model training while ensuring reliability guarantees.

Consider a deep neural network \(f(x)\) with \(L\) hidden layers and a sequence of widths \(\{ d_k \}_{ k=0}^{L+1}\) using ReLU activation. The network \(f(x)\) can be represented as:
\begin{equation}
f(x) = \omega_{L+1} \sigma( \omega_{L}  \ldots \sigma( \omega_2 \sigma(\omega_1 x + v_1 ) + v_2)\ldots +v_L),
\end{equation}
where \(\sigma(u) = \max(u,0)\) denotes the ReLU activation function, \(\omega_l\) is the weight matrix for layer \(l\), and \(v_l\) is the corresponding bias vector. 

Assuming the network weights are uniformly bounded by one (a standard simplification), the weights \(w\) are limited to be sparse. We denote the family of sparse neural networks with ReLU activation as \(F(L, \{d_k\}_{k=0}^{L+1}, s)\). Thus, \(F\) is defined as:
\begin{equation}
F = \{ f \colon S \times A \rightarrow \mathbb{R}   \colon   f(\cdot, a) \in F(L, \{ d_k\}_{k=0}^{L+1} , s ) \}.
\end{equation}
Similarly, let \(G( \{ p_j, t_j , \beta_j , H_j \}_{j \in[q]} )\) be the set of compositions of H\"older smooth functions defined on a subset \(S\subseteq \mathbb{R}^r\). These functions facilitate the demonstration of continuity and smoothness in neural networks with ReLU. Defining \(q \in \mathbb{N}\), the class \(G\) is:
\begin{equation}
G= \{ f\colon S \times A \rightarrow \mathbb{R}  \colon f(\cdot, a) \in G( \{ p_j, t_j , \beta_j ,  H_j \} ) \}.
\end{equation}

Specifically, the set \(F\) comprises ReLU networks commonly used in Q-networks, while \(G\) covers a wide range of smooth functions on \(S\times A\). To proceed with further analysis, we introduce two assumptions:

\textbf{Assumption 1:} For any \(f \in F\), it's assumed that \(Of \in G\), where \(O\) is the Bellman optimality operator. This means the composition \((Of)(s,a)\) can be expressed as a composition of H\"older smooth functions.

\textbf{Assumption 2:} With probability measures \(\nu_1\) and \(\nu_2\) on \(S \times A\) that are absolutely continuous with respect to the Lebesgue measure, and a sequence of policies \({ \pi_t }_{t\geq 1}\), we define the \(m\)-th concentration coefficient as:
\begin{equation}
    \kappa(m; \nu_1, \nu_2) = \sup_{\pi_1, \ldots, \pi_m } \biggl [\mathbb{E} _{u_2} \biggl | \frac{ d ( P^{\pi_m}  \cdots P^{\pi_1}\nu_1)   } { d \nu_2} \biggr | ^2 \biggr ] ^{1/2}.
\end{equation}
This coefficient measures the similarity between \(\nu_1\) and \(\nu_2\) based on action sequences from policies \(\pi_1, \ldots, \pi_m\), and is applicable to a broad class of MDPs.

\begin{theorem}\label{theo}
% \textcolor{blue}{This is the main conclusion that we used from the referenced paper, which define the convergence upper bound of DQN. In this upper bound, we can see that reward, action, state will not affect the performance, and our experiments support our conclusion here.}
Let $F$ and $G $ be given in Eq. (21)-(22), with $\{ H_j\}_{j\in[q]}$ being absolute constants. 
For any $j \in [q-1]$, we define $\beta_j^* = \beta_j \cdot \prod_{\ell = j+1} \min ( \beta_{\ell}, 1)$, $\beta_q^* = 1$, and $
\alpha^* =  \max_{ j \in [q] } t_j / (2 \beta_j^* + t_j )$.
For the parameters of $G$,  the sample size $n$ is sufficiently large such that there exists a constant $\xi>0$ satisfying 

\begin{equation}
\begin{aligned}
\max \bigg\{ &\sum_{j=1}^q    (t_j + \beta_j + 1) ^{3 + t_j}, \\
&\sum_{j \in [q] } \log ( t_j + \beta_j)    , \max_{j \in [q] } p_j \bigg\} \lesssim  (\log n )^{\xi}. 
\end{aligned}
\end{equation}

For any $K \in \mathbb{N}$, let   $Q^{\pi_K}$ be the action-value function corresponding to policy $\pi_K$, which is returned based on function class $F$. Then, there exists a constant $C  > 0 $  such that
\begin{equation}
\begin{aligned}
&\|  Q^* - Q^{ \pi_{K}  }\|_{1, \mu}  \leq  C  \cdot \frac{ \phi_{\mu, \sigma} \cdot \gamma} {(1 -\gamma)^2 } \cdot  | A| \cdot   \\
&( \log n )^{1 + 2\xi^* } \cdot n^{(\alpha^* - 1)/2 }+ \frac{4   \gamma^{K+1}   }{(1- \gamma) ^2 }  \cdot R_{\max}.
\end{aligned}
\end{equation}
\end{theorem}

% (\hl{proof of Theorem 1?})

Theorem \ref{theo} provides a crucial insight into the convergence rates of the action-value function in DQN-based data-driven frameworks. Essentially, this theorem demonstrates that under specific regularity conditions, the action-value function approximated by a sparse ReLU network-based function will converge to the optimal action-value function.
The proof sketch follows a similar structure to the details provided in~\cite{fan2020theoretical}, but is omitted here due to space constraints.
Based on the above theorem, we conclude the upper bound of the vanilla data-driven model, where the designed reliability modules are linear equations that will not affect the upper bound of the RL model, thereby achieving the same convergence rate as the data-driven model without any intervention modules. For instance, when we consider the rewards in the RL model, incorporating our  reward intervention module alters the expected reward to become:

\begin{equation}
    r(s,a)' = r(s,a) - |\frac{\phi}{N} \sum_{n} (\hat{b} - \min_{y} b_y)|.
\end{equation}

\noindent where $b$ is BS load. Notably, the addition of this reliability reward module has no impact on the convergence bound as compared to Eq.~(18). In this regard, we demonstrate that our integrated intervention modules have no negative impact on the data-driven optimization process and model training. For a more comprehensive analysis of their positive impact, the following evaluation section will delve into multi-dimension performance validation within the wireless networks.

\section{Implementation and Evaluation Results} \label{sec:exp}

\subsection{Implementation and Experiment Setup}

In this section, extensive experiments are conducted to evaluate the performance of the D-REC framework, aiming to achieve reliable edge caching in wireless networks.

\subsubsection{Evaluation Environment and Data Forecasting}

We consider a cellular wireless network consisting of five BSs. Each BS is equipped with a local cache unit with a capability of 150 cache slots. A BS can only access its local cache unit and is forbidden to retrieve any cache units from its neighboring BSs. Each BS provides service for eight clients. The service can be overlapped, e.g., one client can be served by up to two BSs. In this case, the cache decision is determined based on the current loads of both BSs.

For the reliability intervention modules, DT and state modules (see Fig.~\ref{fig: saferl}(a)) are integrated into D-REC by default. The DT module generates user requests and the state module extends the state space. In the DT, the frequency of content occurrence distribution follows the Zipf distribution with \( p = 0.8\) by default. 
In the V-H twinning process of DT, we set the learning rate to 0.1, batch size to 64, and asynchronous update threshold to 0.01. In the RL optimization process, we set the learning rate to 0.1, the batch size to 64, and the reward discount factor $\gamma$ to 0.95. The unreliable action mutation threshold $L$ in the action intervention module is set to 0.2.

\subsubsection{Evaluation Metric}

We evaluate the performance of our proposed D-REC schemes using three main metrics:

\begin{itemize}[leftmargin=*]\itemsep=0pt
    \item \textbf{Cache Hit Rate:} This metric measures the percentage of requests that the corresponding content has been stored in the cache unit at the time of request.
    A higher cache hit rate indicates effective management of the cache replacement problem by the optimization process.

    \item \textbf{Action Mutation Count:} It evaluates the number of action mutations to avoid invalid replacements, reflecting the effectiveness of the action intervention module in D-REC. This metric only works for D-REC schemes when equipped with the action module (see Fig.~\ref{fig: saferl}(b)), since other intervention modules do not perform unreliable action mutations. 
    Reduced mutations indicate the reliability of D-SEC optimization and decrease its dependence on external interventions from the action module.

    \item \textbf{BS Load:} This metric is employed to evaluate the network risks arising from imbalanced traffic load and BS overload. BS load refers to the level of resource usage and demand on a particular BS within a wireless network. It is a measure of the traffic or user requests managed by a specific BS at any given time, typically expressed as a percentage of the BS's capacity or resource utilization. For example, an ideal BS load with 5 BSs would be approximately 20\%.
\end{itemize}

\subsubsection{Baseline Methods}

In our evaluation, we compare our proposed optimization method with several baseline caching strategies, which are listed as follows:

\begin{itemize}[leftmargin=*]\itemsep=0pt
    \item \textbf{Basic DQN}~\cite{wang_drlcache_2023}: This approach primarily utilizes a DQN for making cache-replacement decisions. The DQN model is adept at predicting the most suitable content to cache, taking into account the prevailing network conditions and user requests. Nonetheless, this method differs from ours in a significant aspect -- it does not incorporate any intervention module to ensure the reliability of cache decisions. REC is a modified version of Basic DQN that integrates with the state module only.

    \item \textbf{Random Caching}~\cite{7011389}: This strategy randomly selects cache slots for replacement without considering the patterns of user requests or content popularity. This naive approach does not adapt to changing network conditions or user behavior.

    \item \textbf{Least Recently Used (LRU)}~\cite{10.1145/170035.170081}: 
    This widely-used caching strategy replaces the least recently accessed content, based on the assumption that items not used recently are less likely to be needed in the future. 

    \item \textbf{Least Frequently Used (LFU)}~\cite{970573}: 
    LFU operates by evicting the least accessed content. It meticulously records the access frequency of each data item, ensuring that those accessed more often are given priority for retention. % 

    \item \textbf{Most Frequently Used (MFU)}~\cite{10.5555/1286760.1286772}: 
    MFU operates contrary to the LFU, assuming that content accessed frequently in the past is unnecessary in the future. It prioritizes the replacement of the most frequently accessed data content.
\end{itemize}

\subsection{Evaluation Results}

\begin{figure}[ht]
	\centering
        \includegraphics[scale = 0.33]{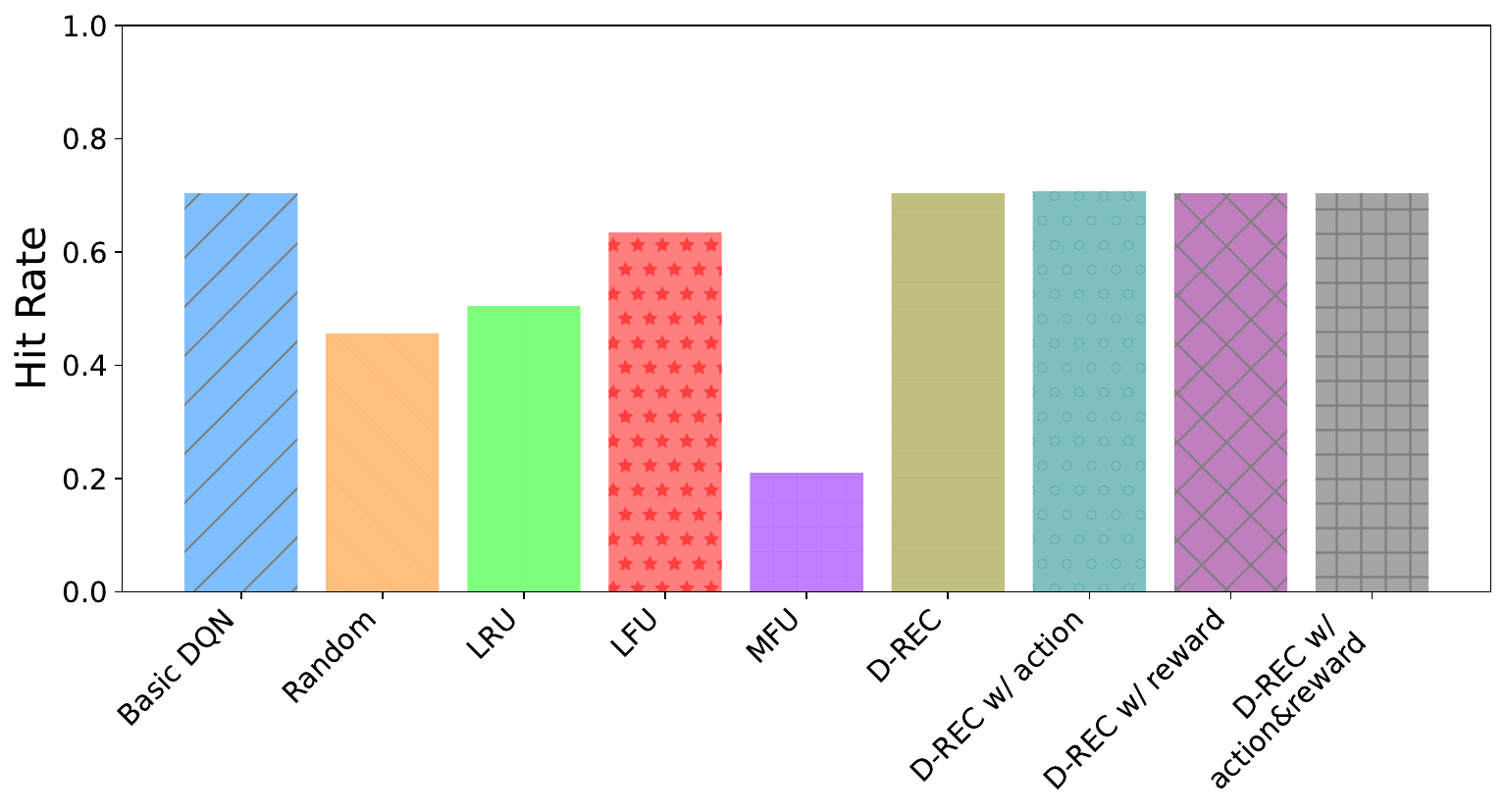}
	\caption{Cache hit rate when incorporating reliability intervention modules.}
	\label{fig: hit_rate_large}
		\vspace{-.1in}
\end{figure}

\subsubsection{D-REC achieves higher cache hit rate}
We first evaluate cache hit rates under different caching policies. 
Figure~\ref{fig: hit_rate_large} demonstrates that D-REC outperforms other baseline algorithms with its noticeably higher cache hit rate and rapid convergence. In contrast, other algorithms converge significantly slower and achieve lower cache hit rates. D-REC excels in its ability to adapt to environment dynamics. This is achieved by training the RL-based optimization model under varying user requests generated from the network DT.
The high cache hit rate and fast convergence rate of D-REC make it a compelling backbone for our reliability interventions.

\subsubsection{Reliability modules have no negative effect on caching optimization}
The encouraging outcomes from our experiments have led us to incorporate the reliability modules detailed in Sec~\ref{sec:modules} into our D-REC optimization, thereby bolstering network stability. 
As depicted in Fig.~\ref{fig: hit_rate_large}, the integration of these modules to D-REC does not adversely affect the cache hit rate. These results align with our theoretical predictions in Sec.~\ref{sec:theoretical_results}, confirming that the integration of D-REC modules does not diminish the network performance.

\begin{figure}[ht]
	\centering
	\includegraphics[scale = 0.4]{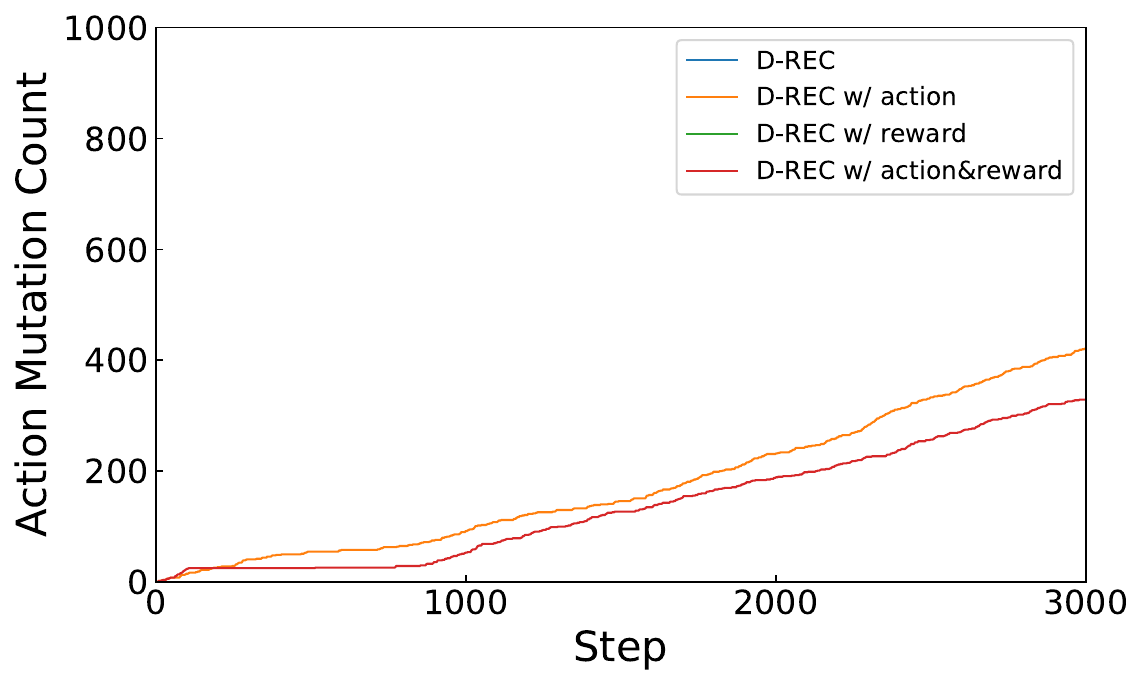}
	\caption{Unreliable action mutation count with different reliability modules. Note that the pure D-REC and D-REC with the reward intervention module do not perform any action mutations, resulting in a count that remains at 0.}
	\label{fig: action_count_modules}
		\vspace{-.1in}
\end{figure}
\begin{figure*}[ht]
	\centering
	\includegraphics[scale = 0.3]{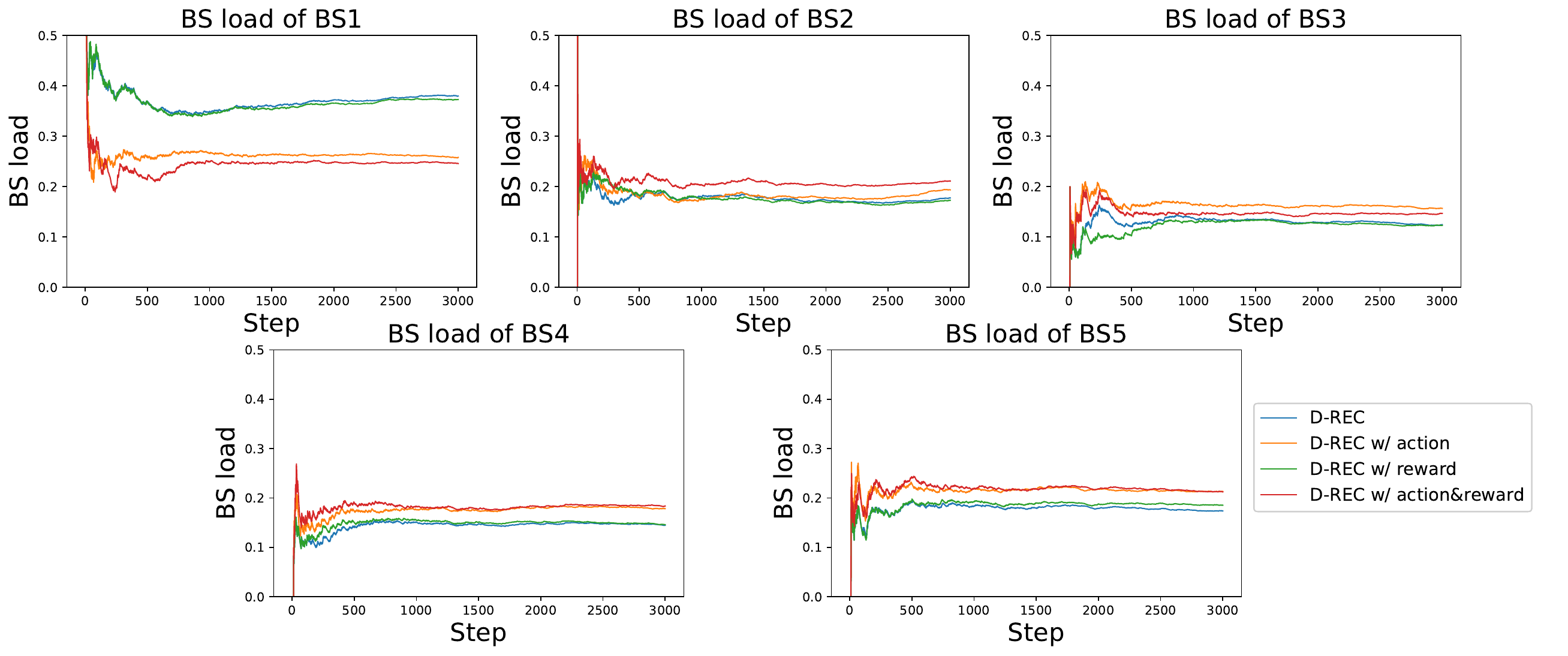}
	\caption{BS load with incorporating different reliability intervention modules.}
	\label{fig: balance}
		\vspace{-.1in}
\end{figure*}
\begin{figure*}[ht]
	\centering
	\includegraphics[scale = 0.65]{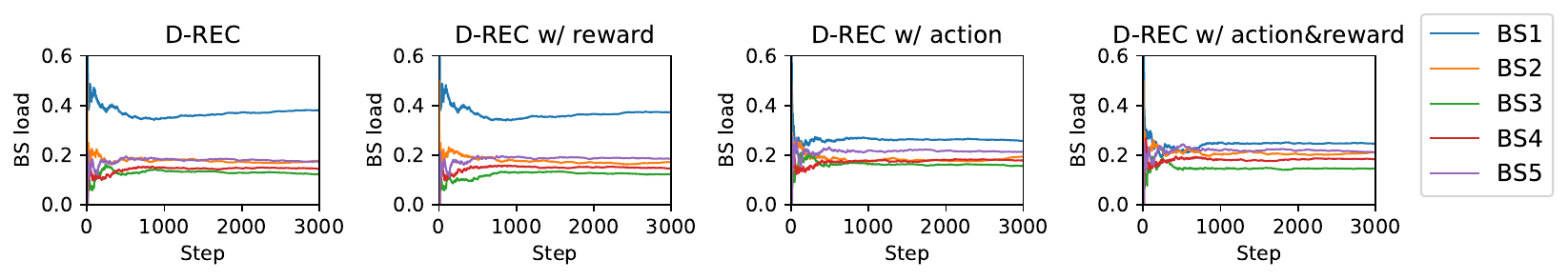}
	\caption{Traffic load across different base stations.}
	\label{fig: overload}
		\vspace{-.1in}
\end{figure*}

\subsubsection{Reliable action module significantly improves load balance between BSs}
Our analysis of D-REC, integrated with a reliable action module, is depicted in Fig.~\ref{fig: action_count_modules}, illustrating the correlation between the number of action mutations and running steps. 
This highlights the occurrence of unreliable actions, underscoring our proposed approach's ability to detect and rectify them through mutation. 
A lower mutation count indicates a reduction in any unreliable actions, crucial for mitigating network risks like BS overload issues. Notably, the mutation count for both pure D-REC and D-REC with the reward intervention module remains zero, as they do not modify unreliable actions during the optimization process. However, D-REC equipped solely with the action module demonstrates a substantially higher mutation rate -- 74\% greater at the 1000$^{\rm th}$ step -- compared to when both action and reward intervention modules are implemented. As Fig.~\ref{fig: saferl}(c) highlights, the reward module significantly influences the RL model during training, effectively reducing action mutations and thereby enhancing network reliability with fewer unreliable decisions.

Next, we delve into how D-REC deals with the network risk arising from the load imbalance and the BS overload. Fig.~\ref{fig: balance} measures the BS load among the five BSs. In an ideal load-balancing case, the normalized load of each BS is expected to be close to 0.2. In Fig.~\ref{fig: balance}, we observe that D-REC without reliability intervention modules has the most unbalanced BS load, while D-REC with both action and reward modules can achieve near-optimal load balance through the strategic mutations. 
In terms of the BS load performance of BS$_1$, D-REC with only the reward intervention module shows a 1\% performance improvement, while D-REC with either the action module or both modules results in a substantial 31\% to 39\% improvement. To provide further clarity, Fig.~\ref{fig: overload} demonstrates that D-REC with the reward module exhibits improved load balance, but it is not as effective as the configuration with only the action module or both modules. In particular, integrating both action and reward modules balances the load performance significantly, maintaining an almost ideal traffic load among deployed BSs over time. 
On the other hand, by adding reliability intervention modules, we observe that D-REC can converge at a reasonable pace, which shows a similar convergence rate to D-REC without any intervention modules. This observation further aligns with our theoretical analysis as derived in Sec.~\ref{sec:theoretical_results}.

\begin{figure}[ht]
	\centering
	\includegraphics[scale = 0.36]{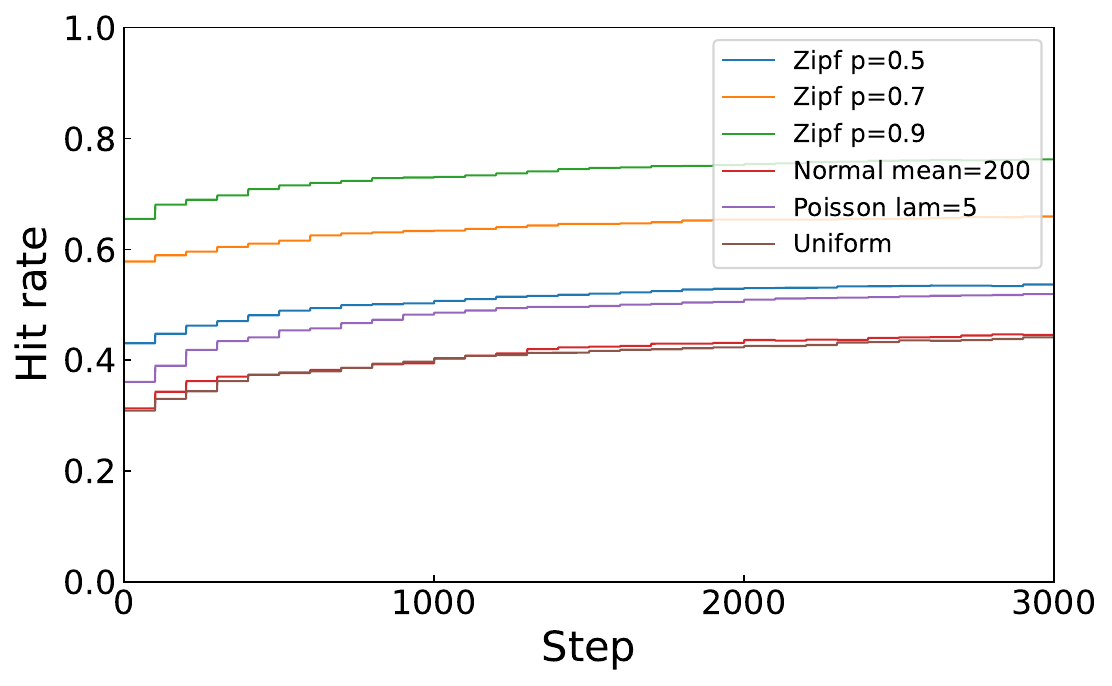}
	\caption{Cache hit rate with the D-REC optimization.}
	\label{fig: hit_rate_datasets}
		\vspace{-.1in}
\end{figure}
\begin{figure}[ht]
	\centering
	\includegraphics[scale = 0.36]{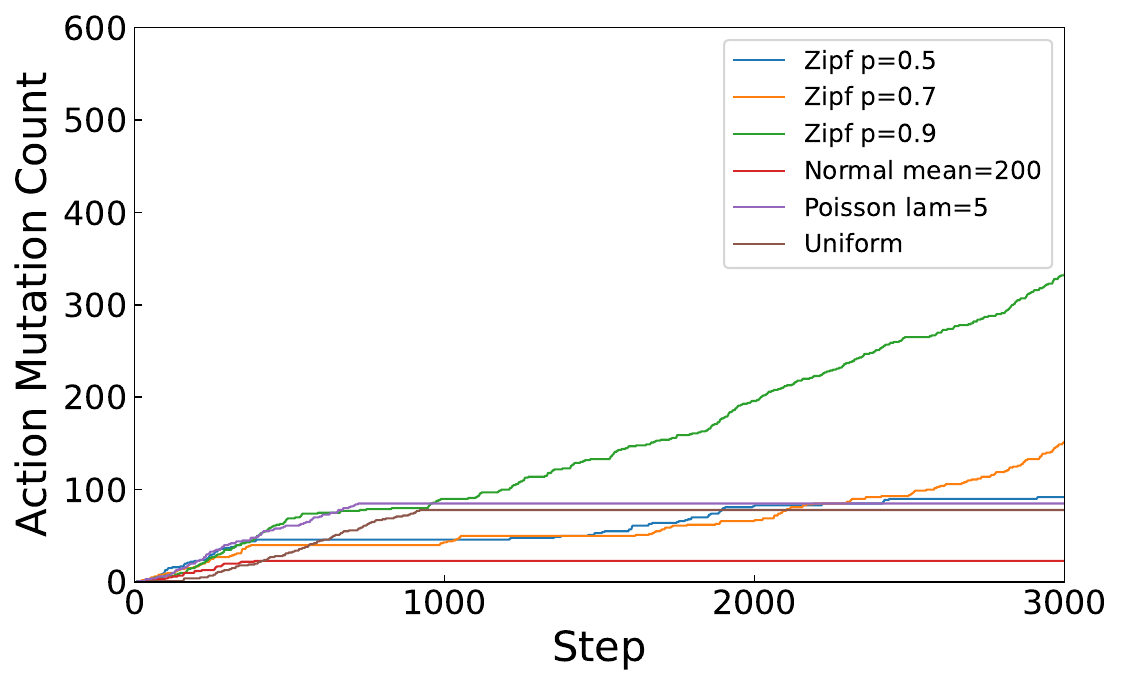}
	\caption{Action mutation count of different datasets using D-REC with both action and reward intervention modules.}
	\label{fig: action_count_dataset}
		\vspace{-.1in}
\end{figure}
\begin{figure*}[ht]
	\centering
	\includegraphics[scale = 0.3]{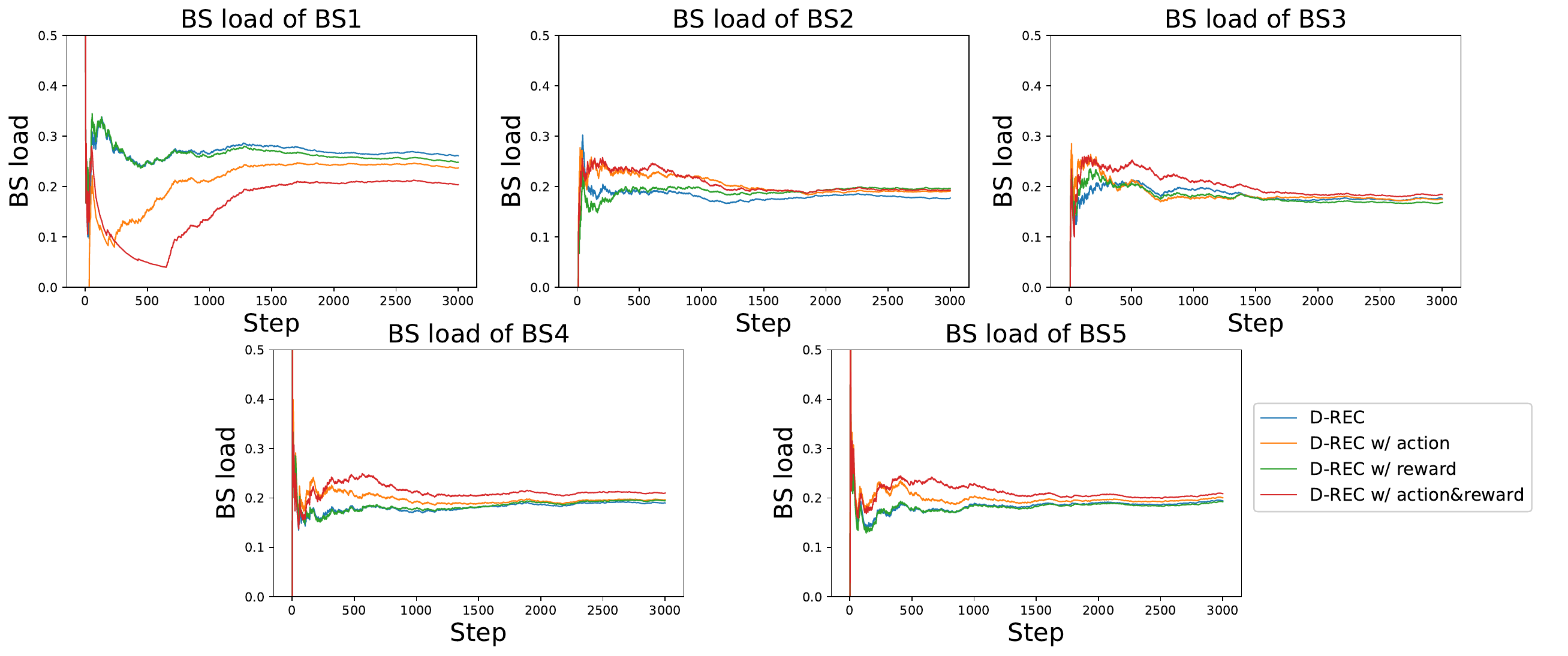}
	\caption{BS load with incorporating different reliability intervention modules under the normal distribution of client requests.}
	\label{fig: balance_dt}
		\vspace{-.1in}
\end{figure*}
\begin{figure*}[ht]
	\centering
	\includegraphics[scale = 0.65]{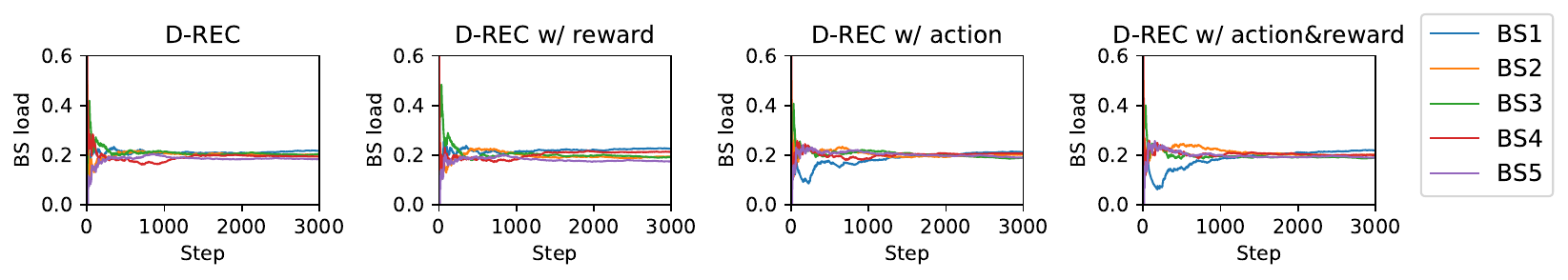}
 \vspace{-0.2cm}
	\caption{Traffic load across different base stations.}
	\label{fig: overload_dt}
		\vspace{-.1in}
\end{figure*}

\begin{figure}[ht]
	\centering
	\includegraphics[scale = 0.35]{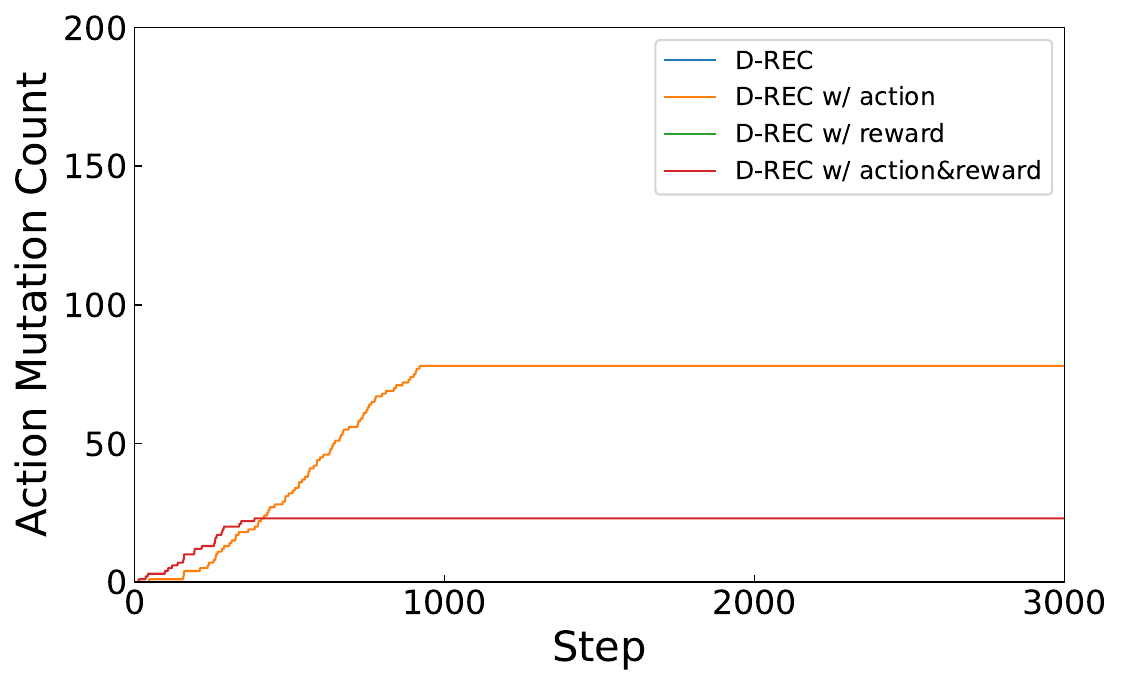}
 \vspace{-0.2cm}
	\caption{The action mutation count of different approaches. The D-REC and ``D-REC w/ reward'' do not perform any action mutation.}
 \vspace{-0.0cm}
	\label{fig: action_count_dt}
		\vspace{-0.2cm}
\end{figure}

\vspace{+0.3cm}
\subsubsection{Network digital twin enhances the stability of D-REC optimization under diverse wireless caching scenarios}

DTs play a pivotal role in enhancing the distribution of client requests within the D-REC framework, specifically by modeling the frequency distribution of client request occurrences over time intervals. The validation of algorithms on DTs before their direct deployment in real-world networks significantly mitigates the risk of system instability.

In the preceding subsection, we intentionally varied the shape parameter $p$ of the Zipf distribution, exploring values from 0.5, 0.7, to 0.9. We also evaluated client requests using normal, uniform, and Poisson distributions to evaluate D-REC's capability to maintain system stability across diverse network scenarios. Notably, parameters such as the mean value and standard deviation for the normal distribution, as well as the predefined expected rate of occurrences in the Poisson distribution, were meticulously set to ensure comprehensive evaluation.
Validation of D-REC's performance under different client request distributions is presented in~Fig.~\ref{fig: hit_rate_datasets} and Fig.~\ref{fig: action_count_dataset}, focusing on cache hit rates and action mutation counts, respectively. D-REC achieves its highest cache hit rate, outperforming the normal distribution by 28\% when employing a Zipf distribution with $p=0.9$. However, it's essential to note that the Zipf distribution leads to more frequent action interventions compared to the normal distribution. D-REC's adaptability to diverse and rare caching scenarios contributes to a more balanced network through this training process.

Furthermore, we validate D-REC's optimization performance in a normal distribution context using data from DTs, particularly focusing on rare scenarios of user requests. As demonstrated in Fig.~\ref{fig: balance_dt}, the pure REC exhibits significant load imbalance among BSs. In contrast, the inclusion of both action and reward modules results in the optimal load balance. Fig.~\ref{fig: overload_dt} illustrated that D-REC, equipped with both reward and action modules, effectively mitigated the overload issue. This configuration outperforms setups where only one of the modules is utilized. 
In addition, as shown in Fig. \ref{fig: action_count_dt}, the action mutation count can be significantly reduced, particularly when compared to the network scenario with the Zipf traffic distribution (as shown in Fig.~\ref{fig: action_count_modules}). This underscores the versatility of the network DT, which can be leveraged to validate models in both common and rare scenarios before actual network deployment.

\begin{table}[ht]
\centering
\caption{Average hit rate among all BSs under different DT module configurations}
% \scriptsize
\begin{tabular}{|c|c|}
\hline
\textbf{Configuration} & \textbf{Hit Rate} \\ \hline
REC & 0.833 \\ \hline
D-REC & 0.878 \\ \hline
REC w/ action & 0.846 \\ \hline
D-REC w/ action & 0.878 \\ \hline
REC w/ reward & 0.839 \\ \hline
D-REC w/ reward & 0.880 \\ \hline
REC w/ action \& reward & 0.838 \\ \hline
D-REC w/ action \& reward & 0.881 \\ \hline
\end{tabular}
\label{tab: dt_hr}
\end{table}

\begin{figure}[ht]
	\centering
	\includegraphics[scale = 0.43]{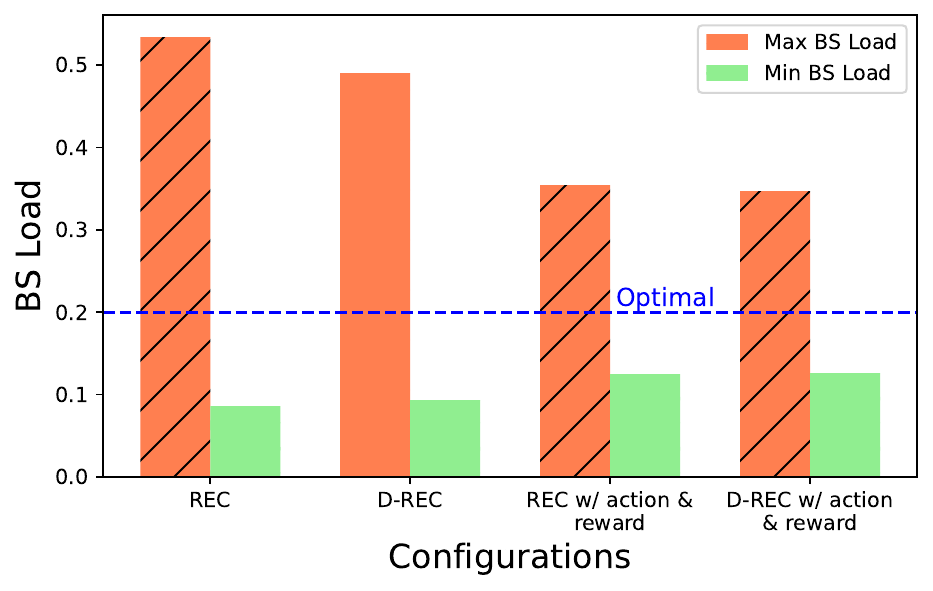}
 \vspace{-0.2cm}
	\caption{BS load under different DT module configurations.}
	\label{fig: dt_load}
		\vspace{-.1in}
\end{figure}

\subsubsection{Network digital twin enhances caching performance and decision-making in real-world scenarios}

To evaluate the effectiveness of our reliability designs, particularly the integration of DT modules, we employed real-world cache request traces from Twitter's Memcached production environment. Memcached is a widely-used in-memory key-value store designed for caching data to alleviate database load and enhance application performance. The dataset encompasses access logs over 7 days, providing insights into the dynamic nature of cache usage in the Twitter platform. Each log entry details key access information, including the operation type (e.g., get, set, delete) and the size of the associated values.
For our reliable caching configurations that incorporate the DT module, we initially train the model in the scenarios generated by our constructed NDT system before applying it to the real-world dataset. Conversely, the configurations without DT modules are trained directly on the real-world dataset. As a comparison point, REC in this evaluation case refers to a basic DQN with only the state intervention module, without the integration of the DT module.
Table~\ref{tab: dt_hr} shows the substantial impact of integrating the DT module into the D-REC framework for wireless caching optimization. The addition of the DT module to the base D-REC configuration resulted in an increase in the average cache hit rate from 0.833 to 0.878, indicating enhanced cache utilization efficiency. Besides, when the DT module is combined with the action and reward intervention modules, the average hit rate can reach its peak at 0.881, underscoring the effectiveness in refining the optimization process.
Furthermore, as depicted in Fig.~\ref{fig: dt_load}, the incorporation of the DT module leads to a more balanced BS load distribution in the network. The maximum BS load decreases from 0.534 in the basic REC configuration to 0.490 with the DT module and further reduces to 0.347 when including all designed intervention modules. This decline in maximum BS load indicates a more equitable allocation of network resources, resulting in enhanced sustainability of network operations. 
Additionally, the minimum BS load also raises from 0.086 to 0.126 with all intervention modules in the framework, emphasizing the role of reliability modules in fostering a more uniform distribution of network load.

%% !TEX root = main.tex

\vspace{-0.2cm}
\section{Conclusion} \label{sec:conclusion}

This work presents D-REC, an innovative optimization framework that merges reliable RL with digital twins to boost edge caching performance in nextG wireless networks. It tackles the crucial problem of overlooking reliability considerations in current data-driven optimization approaches, which can potentially cause BS overloads, imbalances, and diminished user experiences. By incorporating strategic intervention modules into the CMDP process, D-REC ensures adaptive modifications to the modules of actions, rewards, and states, aligning them with reliability constraints to minimize the likelihood of network failures. 
Our theoretical analysis shows that such reliability designs do not affect the convergence rate of the RL optimization process. 
Additionally, 
the inclusion of DTs as RL optimizers and safeguards in D-REC allows the utilization of diverse data patterns for predictive evaluation of cache replacement policies. This enhances the network's adaptability and efficacy in managing cached content across densely deployed small-cell BSs. Comprehensive experiments confirm that D-REC outperforms traditional approaches in cache hit rate, and load balance, and effectively enforces a reliable optimization process.

% \section*{References}

% \newpage

\bibliographystyle{IEEEtran}
\bibliography{refs}

% \newpage
% \input{response}
\section{Biography Section}

\vspace{-0.5in}
\begin{IEEEbiography}
    [{\includegraphics[width=1in,height=1.25in,clip,keepaspectratio]{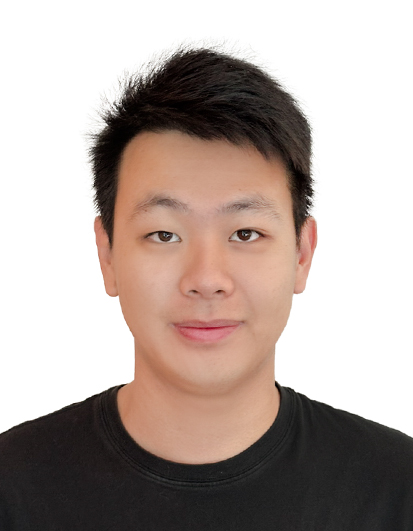}}]{Zifan Zhang}
is a Ph.D. student with the Department of Computer Science at North Carolina State University, USA.
He received his Bachelor's and Master's degrees in Electrical and Computer Engineering at the Ohio State University, USA, in 2021 and 2023, respectively. His current research interests include wireless networking, digital twins, distributed learning, and model security.
\end{IEEEbiography}

% \vspace{-0.5in}
\begin{IEEEbiography}
    [{\includegraphics[width=1in,height=1.25in,clip,keepaspectratio]{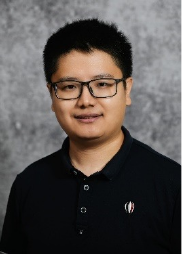}}]{Yuchen Liu}
is currently an Assistant Professor with the Department of Computer Science at North Carolina State University, USA. He got his Ph.D. degree at the Georgia Institute of Technology, USA. His research interests include wireless networking, generative AI, distributed learning, mobile computing, and digital twins. He has received several best paper awards at IEEE and ACM conferences. He currently serves as Associate Editors of IEEE Transactions on Green Communications and Networking, IEEE Transactions on Machine Learning in Communications and Networking, and Elsevier Computer Networks.
\end{IEEEbiography}

\vspace{-0.5in}
\begin{IEEEbiography}
    [{\includegraphics[width=1in,height=1.25in,clip,keepaspectratio]{bio/peng.pdf}}]{Zhiyuan Peng}
received the B.Eng. degree in electronics and information engineering from Harbin Institute of Technology, China, in 2017, and the Ph.D. degree in electronic engineering from The Chinese University of Hong Kong, Hong Kong, in 2023. He is currently a Post-doc with the Department of Computer Science at North Carolina State University, USA. His research interests include wireless networking, Bayesian optimization, generative AI, and speech and language processing.
\end{IEEEbiography}

\vspace{-0.5in}
\begin{IEEEbiography}
    [{\includegraphics[width=1in,height=1.25in,clip,keepaspectratio]{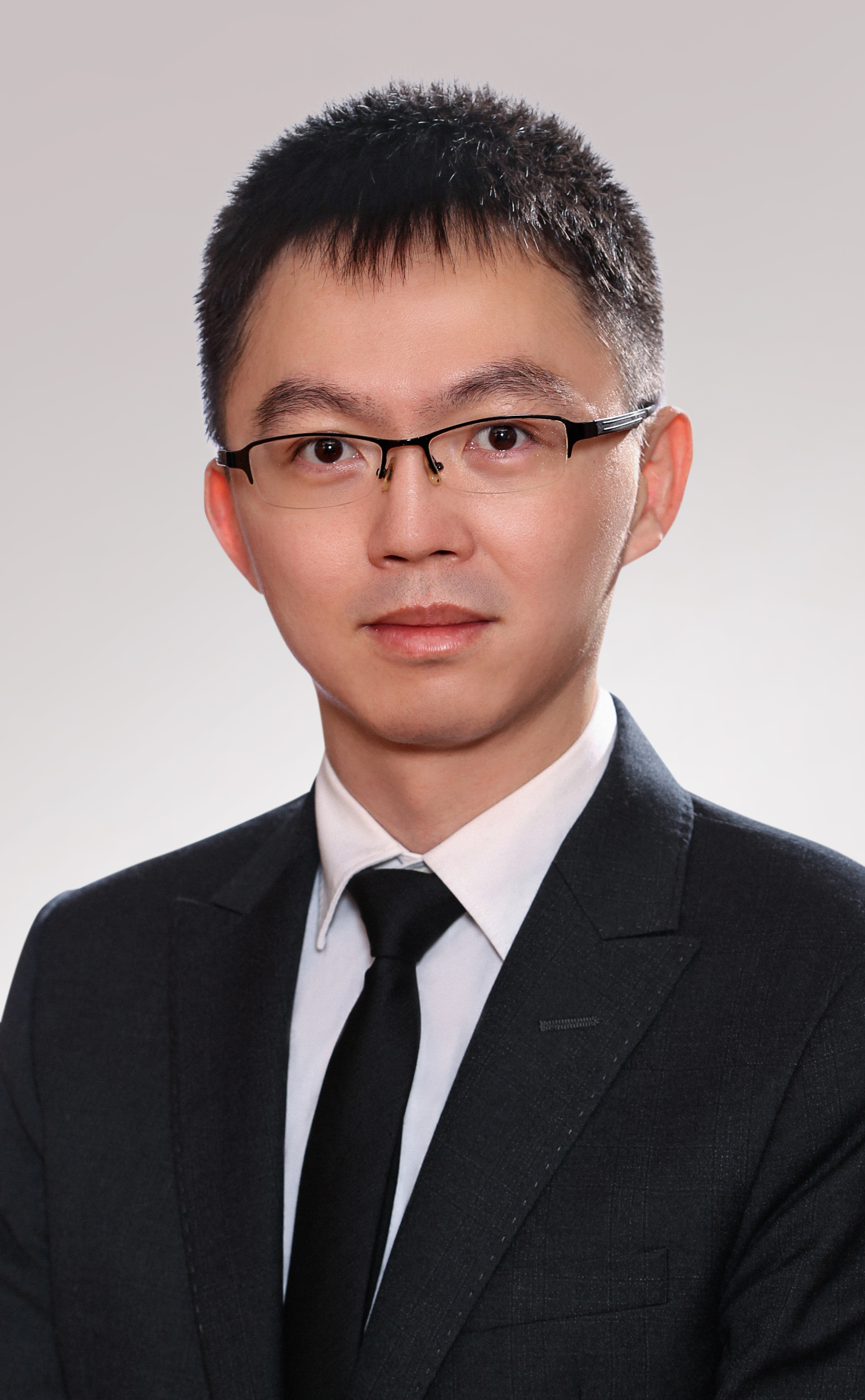}}]{Mingzhe Chen}
is currently an Assistant Professor with the Department of Electrical and Computer Engineering and Frost Institute of Data Science and Computing at University of Miami. His research interests include federated learning, reinforcement learning, virtual reality, unmanned aerial vehicles, and Internet of Things. He has received four IEEE Communication Society journal paper awards and four conference best paper awards. He currently serves as an Associate Editor of IEEE Transactions on Mobile Computing, IEEE Transactions on Communications, IEEE Wireless Communications Letters, IEEE Transactions on Green Communications and Networking, and IEEE Transactions on Machine Learning in Communications and Networking.
\end{IEEEbiography}

\vspace{-0.5in}
\begin{IEEEbiography}
    [{\includegraphics[width=1in,height=1.25in,clip,keepaspectratio]{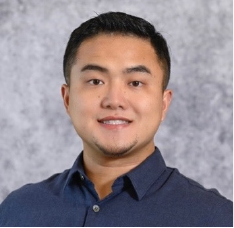}}]{Dongkuan Xu}
is an Assistant Professor in the CS Department at NC State and leads the NCSU Generative Intelligent Computing Lab. His research is fundamentally grounded in advancing Artificial General Intelligence, particularly the automated planning, reliable reasoning, and efficient computing of generative AI systems. He has been honored with the Microsoft Accelerating Foundation Models Research Award 2024, the NCSU Carla Savage Award 2024, and the Best Paper Award of ICCCN 2023. He serves as the Column Editor for the ACM SIGAI Newsletter and will chair the 1st Workshop on Dataset Distillation for Computer Vision at CVPR 2024. 
% In the past, he chaired the 1st Workshop on Deep Learning-Hardware Co-Design for AI Acceleration at AAAI 2023, the Resource Efficient Learning for Knowledge Discovery Workshop at KDD 2023, the session of New Deep Learning Architectures at KDD 2022, and the session of Scalable \& Trustable AI at KDD 2022. 
\end{IEEEbiography}

\vspace{-0.5in}
\begin{IEEEbiography}
    [{\includegraphics[width=1in,height=1.25in,clip,keepaspectratio]{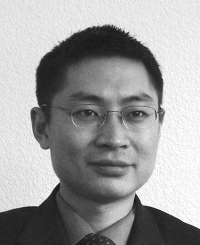}}]{Shuguang Cui}
received his Ph.D in Electrical Engineering from Stanford University, California, USA, in 2005. Afterwards, he has been working as assistant, associate, full, Chair Professor in Electrical and Computer Engineering at the Univ. of Arizona, Texas A\&M University, UC Davis, and CUHK at Shenzhen respectively. His current research interests focus on the merging between AI and communication networks. He was selected as the Thomson Reuters Highly Cited Researcher and listed in the Worlds’ Most Influential Scientific Minds by ScienceWatch in 2014. 
He has also been serving as the area editor for IEEE Signal Processing Magazine, and associate editors for IEEE Transactions on Big Data, IEEE Transactions on Signal Processing, and IEEE Transactions on Wireless Communications, and the Editor-in-Chief for IEEE Transactions on Mobile Computing. 
\end{IEEEbiography}

\end{document}